\documentclass[prb,aps,showpacs,twocolumn,floatfix]{revtex4}
\usepackage{amsfonts}
\usepackage{amssymb}
\usepackage{latexsym}
\usepackage{dcolumn}
\usepackage{epsfig}

\newcommand{\ket}[1]{|#1\rangle}
\newcommand{\bra}[1]{\langle#1|}
\newcommand{\tN}{\text{N}}
\newcommand{\tNe}{\text{N\a{'}eel}}

\begin{document}
\title{N\'eel probability and spin correlations in some nonmagnetic and nondegenerate states
of hexanuclear antiferromagnetic ring Fe$_6$:\\ Application of algebraic combinatorics to
finite Heisenberg spin systems}
\author{Wojciech Florek}
\email{florek@amu.edu.pl}
\author{Sylwia Bucikiewicz}
\email{bucz@spin.amu.edu.pl}
\affiliation{A. Mickiewicz University, Institute of Physics,
ul. Umultowska 85, 61--614 Pozna\'n, Poland}
\date{\today}
\begin{abstract}
 The spin correlations $\omega^z_r$, $r=1,2,3$, and the probability $p_{\tN}$ of finding 
a system in the N\'eel state for
the antiferromagnetic ring Fe$^{\text{III}}_6$ (the so-called `small ferric wheel') are
calculated. States with magnetization $M=0$, total spin $0\le S\le15$ and labeled
by two (out of four) one-dimensional irreducible representations (irreps) of the point 
symmetry group ${\mathrm D}_6$ are taken into account. This choice follows from importance 
of these irreps in analyzing low-lying states in each $S$-multiplet. Taking into account
the Clebsch--Gordan coefficients for coupling total spins of 
sublattices ($S_A=S_B={15\over2}$)
the global N\'eel probability $p^*_{\tN}$ can be determined.  Dependencies of these 
quantities on state energy (per bond and in the units of exchange integral $J$) and the 
total spin $S$ are analyzed. Providing we have determined $p_{\tN}(S)$ etc.\ for other
antiferromagnetic rings (Fe$_{10}$, for instance) we could try to approximate results
for the largest synthesized ferric wheel Fe$_{18}$. Since thermodynamic properties of
Fe$_6$ have been investigated recently, in the present considerations they are not 
discussed, but only used to verify obtained values of eigenenergies. Numerical results
are calculated with high precision using two main tools: (i) thorough analysis of symmetry
properties including methods of algebraic combinatorics and (ii) multiple precision
arithmetic library {\sf GMP}. The system considered yields more than 45 thousands basic
states (the so-called Ising configurations), but application of the method proposed 
reduces this problem to 20-dimensional eigenproblem for the ground state ($S=0$). The 
largest eigenproblem has to be solved for $S=4$; its dimension is 60. These two facts
(high precision and small resultant eigenproblems) confirm efficiency and usefulness
of such an approach, so it is briefly discussed here. 

\end{abstract}


\maketitle

\section{Introduction}

Magnetic hysteresis with a pure molecular origin first time was 
observed \cite{can91,nat93} in the mixed-valence Mn complex
[Mn$_{12}$O$_{12}$(CH$_3$COO$_{16}$(H$_2$O)$_4$]$\cdot$2CH$_3$COOH$\cdot$4H$_2$O
(commonly referred to as Mn$_{12}$acetate),
synthesized and investigated by Lis in 1980.\cite{lis}
Since this discovery mesoscopic magnetic systems, comprising transition-metal
ions, have attracted a great deal of attention. Many large molecular
aggregates have been fabricated and investigated both experimentally and
theoretically (including numerical simulations). This includes
Mn molecules like Mn$_6$,\cite{can88,jpcm01} a ferromagnetic copper 
ring,\cite{las98} mixed Mn-Cr compounds like 
[Cr$^{\text{III}}$(CNMn$^{\text{II}}$L)$_6$](ClO$_4$)$_9$,\cite{scu96}
some nickel complexes,\cite{jpcm01,blake} and many polynuclear iron 
clusters known as Fe$_n$ with $n$ varying from two\cite{fe02} through 
20.\cite{papa92,can99} Some interesting features have been also observed in a crystal made
of nanometric molecules (so-called V$_{15}$)\cite{wern00}
and in a single cobalt nanocluster.\cite{jamet00,jamet01}

Such systems of mesoscopic dimensions containing transition-metal ions and
oxygen are attracting increasing interest for several reasons, including
the possibility of observing quantum phenomena on a macroscopic scale. It may lead
to better understanding of the mechanism associated with the transition from simple 
paramagnetic to bulk magnetism (also low-dimensional in magnetic chains). On the 
other hand, clusters comprising transition elements exist in several 
metalloenzymes and metalloproteins\cite{blon90} as ferritin storing
Fe in mammals.\cite{stpie96,tejada97} Moreover, nanostructures are part of the development
of high density magnetic recording media\cite{jamet01} and quantum 
computing.\cite{sell99}

Molecular magnets show both antiferro- and ferromagnetic behavior, though 
in many cases single spins are coupled antiferromagnetically.\cite{can99}
Incomplete compensation of this coupling may lead to the large spin ground
state of a cluster, e.g.\ $S=10$ in Fe$_8$ and 
Mn$_{12}$,\cite{can99,fe08b,sang97,fe08c,prok98} what  
is related to a strong easy-axis term (Ising-type anisotropy). There are
also purely ferromagnetic molecules as dodecanuclear nickel complex (Ni$_{12}$) 
with the ground state characterized by $S=12$.\cite{blake,can99,jpcm01} 
The other class of nanomagnets comprises molecules where the cancellation is
complete and the ground state has $S=0$. The most profound example is provided
by a series of antiferromagnetic Fe(III) rings Fe$_6$, Fe$_{10}$, Fe$_{12}$, and
Fe$_{18}$.\cite{can99} Recently a new octanuclear cluster Fe$_8$ has been synthesized 
and investigated.\cite{saal97,wald01} A deviation from coplanarity in the case of 
Fe$_{10}$ ring,
about 9.8\,\AA\ in diameter, is less than 0.01\,\AA,\cite{can99,fe06a} so
these rings can be considered as two-dimensional structures of very high symmetry:
at least idealized or approximate point-group symmetry isomorphic to the dihedral 
group D$_n$ (S$_6$ or D$_{5d}$, for example).\cite{fe06b}

Due to the presence of organic ligands wrapping the clusters, the inter-cluster
interactions are vanishingly small (or can be reduced by dissolving the 
clusters in the appropriate solvent).\cite{fe08b} Each magnetic
object has perfectly defined size and new techniques of molecular chemistry 
provide ensembles of iso-oriented nanomagnets.\cite{fe06c,sang97} Properties of such  
systems can be very well simulated using contemporary computers within the
framework of finite spin models.\cite{jpcm01,fe06b} However, direct calculations
of eigenenergies and eigenstates, even using the Lanczos exact diagonalization 
technique, are subjected to the so-called ``combinatorial explosion''. For example,
in the case of the largest ``ferric wheel'' Fe$_{18}$ one has to deal with more
than $10^{14}$ basic states. This is especially important in the case of antiferromagnetic
systems when the ground state lies in the subspace of states with the total magnetization
$M=0$ and determination of some low-lying states only is a formidable task. 
This can be done a bit more manageable exploiting symmetry properties of a
given Hamiltonian $\mathcal{H}$ (including the time-reversal symmetry) and good quantum 
numbers, i.e.\ eigenvalues of operators commuting with $\mathcal{H}$. In the latter
case we consider the total magnetization $M$ (the eigenvalue of $S^z=\sum_{j=1}^n s_j^z$)
and the total spin $S$ (the eigenvalues of the squared total spin 
${\mathbf S}^2=(\sum_{j+1}^n{\mathbf s}_j)^2$ are $S(S+1)$, of course). While model
Hamiltonians commute with $S^z$ as a rule, the additional anisotropy terms break 
commutation with ${\mathbf S}^2$.\cite{fe06c} However, if the isotropic Heisenberg
interactions 
  \begin{equation}\label{ham}
    {\mathcal H}_{\text{H}}=J\sum_{\langle jk\rangle}^n {\mathbf s}_j\cdot{\mathbf s}_k
  \end{equation}
(the sum is taken over the nearest-neighbor pairs) are dominant, then the effect of
magnetic anisotropy and Zeeman interactions can be considered separately for
each $S$-multiplet.\cite{fe06c,bengat} Therefore, in this article we restrict 
ourselves to Hamiltonians commuting with ${\mathbf S}^2$. It should be stressed
that this includes interactions with different ranges (e.g.\ next nearest-neighbors)
and forms (e.g.\ biquadratic terms $({\mathbf s}_j\cdot{\mathbf s_k})^2$).

As regards symmetry properties, the simplest task is to take into account the
time-reversal symmetry. It is enough to consider states with $M\ge0$ since the others
(with $M<0$) are determined in a straightforward way. 
The largest eigenproblem dimension is reached considering the subspace 
$L_0$ comprising states with $M=0$ (the are more 
than $5.5\cdot10^{12}$ such basic states for $s={5\over2}$ and $n=18$). The point-group
symmetry $G$ enters the problem considered via the Schur Lemma: eigenspaces of operators
commuting with all representation operators $P(g)$, $g\in G$, are labeled by 
the irreducible representations (irreps) of $G$. However, numbers $n(\Gamma)$ in
the decomposition
 \begin{equation}\label{decom}
      P =\bigoplus_\Gamma n(\Gamma)\Gamma
 \end{equation}  
 are very large, even if $P$ is restricted to $L_0$. Let $L^{\Gamma r}$, 
$1\le r\le n(\Gamma)$, be copies of the eigenspace of an irrep $\Gamma$ spanned
over vectors $\{\ket{\Gamma\gamma}\mid 1\le\gamma\le\dim L^\Gamma\}$. Solving an 
eigenproblem for any operator of symmetry $G$ we have to construct an
$n(\Gamma)\times n(\Gamma)$ matrix for an arbitrarily chosen, but fixed, vector
index $\gamma$. To do it in an efficient way, the single, meaningless, repetition 
index $r$ has to be replaced by a series of indices with well-determined mathematical
notion. In the case considered the representation $P$ is in fact a permutation
representation,\cite{jmmm99,cpc01,ker} so the methods of algebraic combinatorics
give an efficient way to deal with this problem. Such approach to finite spin
systems has been discussed recently in a series of papers.\cite{app1,app2}
In this work we aim to present and discuss results obtained for the 
``small ferric wheel'' Fe$_6$. The reason of such a choice is three-fold: (i)
this molecule has been well investigated,\cite{fe06a,fe06b,fe06c,fe06d,las98} so
we can verify our results; (ii) the number of states are relatively small
($6^6=46\,656$ and $\dim L_0=4332$), then many characteristics can be obtained
without tedious calculations; (iii) all numerical algorithms and procedures can
be tested before going to larger problems.

The paper is organized as follows. In Sec.~\ref{fering} a short account
of the iron antiferromagnetic rings is presented. Combinatorial
classification of finite spin system states is briefly discussed in Sec.~\ref{class}.
Section~\ref{fe6-eigen} contains eigenproblem solutions 
for Fe$_6$ cluster. Eigenstates obtained are analyzed in Sec.~\ref{fe6-res} (e.g.\
static spin correlations are calculated). Some concluding remarks are made in 
Sec.~\ref{diss}. Due to reasons given in Sec.~\ref{fe6-eigen} states corresponding to 
two one-dimensional irreps are considered only. Hence, the energy spectrum
is not determined and the thermodynamic quantities are not calculated.

\section{Antiferromagnetic ferric wheels}\label{fering}

There are several mesoscopic clusters comprising iron ions bridged by oxo-groups with
different magnetic properties.
Some of them have the ground state with high total spin $S$, e.g.\
Fe$_4^{\text{III}}$(OCH$_3$)$_6$(dpm)$_6$ with $S=5$,\cite{fe04x}  
[Fe$_8^{\text{III}}$(tacn)O$_2$(OH)$_{12}$]Br$_8$ with $S=10$,\cite{fe08a,fe08b} 
and ferromagnetic Fe$_4^{\text{II}}$ with $S=8$.\cite{fe04} On the other hand, in
some systems antiferromagnetic coupling results in complete cancellation of local spins
leading to nonmagnetic ($S=0$) ground state; all such molecules contain iron(III) ions
arranged in cycles\cite{can99} or dimers as in [Fe(OMe)(dbm)$_2$]$_2$.\cite{fe02} 
In the first case we have three types of Fe$_6$ rings, 
a medium ferric wheel [Cs$\subset$Fe$_8\{$N(CH$_2$CH$_2$O)$_3\}_8$]Cl,\cite{saal97}
the Fe$_{10}$(OCH)$_3$)$_{20}$(C$_2$H$_2$O$_2$Cl)$_{10}$ 
ring,\cite{fe06b,fe10} and the largest ferric wheel Fe$_{18}$.\cite{can99}
There is also a ring-shaped molecule Fe$_{12}$ with a net spin $S=0$ in the ground state, but
it does not exhibit full cyclic symmetry.\cite{can99} The number of basic states is 
given as $6^n$, $n=2,6,8,10,12,18$, so we deal with eigenproblems of dimensions
36, 46\,656, 1679\,616, 60\,466\,176, 2176\,782\,336, and 101\,559\,956\,668\,416, respectively.
The dimensions of $L_0$, the subspace containing states with $M=0$ are as 
follows: 6, 4332, 135\,954, 4395\,456, 144\,840\,476, and 5542\,414\,273\,884;
the approximate ratios $6^n/\dim L_0$ are 6.00, 10.77, 12.35, 13.76, 15.03, and 18.32,
respectively, so $\dim L_0$ increases a bit slower than $6^n$.

In this work the small ferric wheel is considered; the following three molecules
of this type are known and investigated:\cite{fe06a}
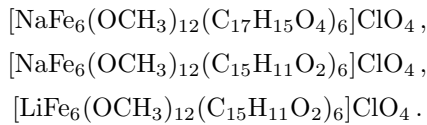
\begin{center}
   [NaFe$_6$(OCH$_3$)$_{12}$(C$_{17}$H$_{15}$O$_4$)$_6$]ClO$_4$\,, \\[5pt]
   [NaFe$_6$(OCH$_3$)$_{12}$(C$_{15}$H$_{11}$O$_2$)$_6$]ClO$_4$\,, \\[5pt]
   [LiFe$_6$(OCH$_3$)$_{12}$(C$_{15}$H$_{11}$O$_2$)$_6$]ClO$_4$\,.
\end{center}
  The first cluster has crystallographically imposed S$_6$ point-group symmetry, whereas
in the other cases this symmetry is approximate.\cite{fe06a} Therefore, the results
obtained here are mostly relevant to the first one.  

\section{Classification of states by means of algebraic combinatorics}\label{class}

The method proposed can be roughly divided into four steps: combinatorial,
linear, magnetic, and numerical. The last one consists in efficient implementation
of proposed algorithms, working-out methods of storing of configuration etc.
These problems have been recently presented elsewhere,\cite{cpc01,cmstm,cmsta}
so they are left out in the present paper. The second and third steps are slightly
modified standard procedures known since the famous paper of Bonner and Fisher\cite{BF64}
and reused lately in many articles, for example by Waldmann in Ref.~\onlinecite{wald00}.
Therefore, below we concentrate on the first step since combinatorial methods
(especially methods of algebraic combinatorics) are not so popular. More detailed
discussion on this topic, from the mathematical point of view, can be found in 
a monograph by Kerber.\cite{ker}

To begin with we have to introduce the notion of a (finite) group action and
define two structures: an orbit and a stabilizer. We say that a group $G$ acts 
on a set $X$ if for each $g\in G$ and $x\in X$ a product $gx\in X$ is uniquely 
determined and for all $g,g'\in G$, $x\in X$
 \[
  (gg')x=g(g'x)\qquad \text{and}\qquad e_Gx=x\,,
 \]
where $e_G$ denotes the unit element in $G$. If both $G$ and $X$ are finite we call
the action finite. An orbit $G(x)\subset X$ of an element $x\in X$ is a set
containing all $x'\in X$ which can be obtained from $x$, i.e.
 \[
   G(x)=\{gx\mid g\in G\}\,.
 \]
 Since an orbit is an equivalence class then it can be represented by any of its
elements. For each $x\in X$ its stabilizer 
 \[
   G_x=\{g\mid gx=x\}\subseteq G
 \]
 contains elements of $G$ leaving $x$ invariant. It can be shown\cite{ker} that 
orders $|G(x)|$ and $|G_x|$ of an orbit and its stabilizer, respectively, are 
related by the following formula
 \[
   |G|=|G(x)|\,|G_x|\,.
 \]
Moreover stabilizers of all elements in a given orbit $G(x)$ are conjugated to each other
(as subgroups of $G$); a class of conjugated subgroups determines a type of an orbit.
For any set $Y$ the action of $G$ on $X$ can be raised to an action of $G$ on functions 
$f\colon X\to Y$ as follows\cite{ker} 
 \begin{equation}\label{mapact}
   \forall\,x\in X\quad (gf)(x) = f(g^{-1}x)\,.
 \end{equation}

In the case of finite systems the so-called Ising configurations (basic states) can be
considered as mappings $\mu\colon X\to Y$, where $X=\{1,2,\dots,n\}$ is a set of node
labels and $Y=\{-s,-s+1,\dots,s\}$ is a set of $z$-projections for the spin number $s$.
Basic states $\ket{m_1,m_2,\dots,m_n}$, $m_j\in Y$ for $1\le j\le n$, are in a 
one-to-one correspondence with functions $\mu$ providing that for all $j\in X$ the 
equality $\mu(j)=m_j$ holds.  The largest (discrete) group
acting on $X$ is the symmetric group $\Sigma_n$ comprising all $n!$ permutations $\sigma$
of the $n$-element set $X$. According with Eq.~(\ref{mapact}) one obtains
 \[
  (\sigma\mu)(j)=\mu(\sigma^{-1}j)\qquad(\forall\;1\le j\le n)
 \]
 or, writing it for basic states $\ket{m_1,m_2,\dots,m_n}$,
 \[
  \sigma\ket{m_1,m_2,\dots,m_n}= 
   \ket{m_{\sigma^{-1}1},m_{\sigma^{-1}2},\dots,m_{\sigma^{-1}n}}\,.
 \]
 Let us consider, for example, a system of $n=4$ spins $s={3\over2}$. One of the Ising 
configurations is a state $\ket{\psi}=\ket{{1\over2},{1\over2},{3\over2},{3\over2}}$. 
Acting with all $4!=24$ permutations on this state one obtains only six different states
(i.e.\ $|G(\ket{\psi})|=6$), 
since a stabilizer $G_\psi$ is given as
 \[
  G_\psi=\{1,(12),(34),(12)(34)\}\,,
 \] 
 where $(jk)$ denotes a transposition of nodes $j$, $k$ and $1$ is the unit element 
(identity) in $\Sigma_n$. To obtain all elements of the orbit $G(\ket{\psi})$ it is 
enough to act with representatives of left cosets $\Sigma_4/G_\psi$, for example
with permutations $1, (13), (14), (23), (24)$, and $(14)(23)$. If $\sigma$ is a 
transposition then $\sigma^{-1}=\sigma$, so 
 \begin{eqnarray*}
  G(\ket{\psi})&=&\{
\ket{{\textstyle {1\over2},{1\over2},{3\over2},{3\over2}}},
\ket{{\textstyle {3\over2},{1\over2},{1\over2},{3\over2}}}, 
\ket{{\textstyle {3\over2},{1\over2},{3\over2},{1\over2}}}, \\ &&\phantom{\{}
\ket{{\textstyle {1\over2},{3\over2},{1\over2},{3\over2}}}, 
\ket{{\textstyle {1\over2},{3\over2},{3\over2},{1\over2}}}, 
\ket{{\textstyle {3\over2},{3\over2},{1\over2},{1\over2}}}
\} \,.
 \end{eqnarray*}
 Note, that this orbit is determined in unambiguous way by a nonordered partition 
$[0,0,2,2]$ of $n=4$ into $2(3/2)+1=4$ non-negative parts. Each entry $k_l$, $l=0,1,\dots,2s$,
of $[k]=[k_0,k_1,\dots,k_{2s}]$ denotes a number of projections $m_j$, $j=1,2,\dots,n$, equal to
$l-s$. Since $[k]$ is a partition of $n$, then $\sum_{l=1}^{2s+1}k_l=n$ and $k_l\ge 0$. 
An orbit containing states with $k_l$ projections $l-s$ is denoted hereafter as
$O[k]$ and is represented, for example, by a configuration 
 \[
  \ket{ \underbrace{-s,-s,\dots,-s}_{k_0\; \mathrm{times}},
    \underbrace{-s+1,\dots,-s+1}_{k_1\; \mathrm{times}},\dots,
    \underbrace{s,s,\dots,s}_{k_{2s}\; \mathrm{times}} }\,.
 \] 
Moreover, a nonordered partition $[k]$ determines the magnetization $M$ for all states
in $O[k]$ since
 \begin{equation}\label{part}
   M=\sum_{l=0}^{2s} (l-s)k_l=\sum_{l=0}^{2s} lk_l -ns\,.
 \end{equation}
 In the example presented one obtains $M=-{1\over2}\cdot0 -{3\over2}\cdot0
 +{1\over2}\cdot2 +{3\over2}\cdot2=4$. The same magnetization $M=4$ is obtained
for states from the four-element orbit $O[0,1,0,3]$ represented by 
a state $\ket{{-{1\over2}},{3\over2},{3\over2},{3\over2}}$. Some group-theoretical
properties of orbits $O[k]$ depend on their type, i.e.\ on a class of conjugated
subgroups of $\Sigma_n$ a stabilizer of states in $O[k]$ belongs to. Such classes are 
represented by the so-called Young subgroups $\Sigma_{[\kappa]}\subset\Sigma_n$,\cite{ker,kerjam}
where $[\kappa]$ is an ordered partition $[\kappa_0,\kappa_1,\dots,\kappa_z]$, $z\le2s+1$, 
of $n$ into no more than $2s+1$ nonzero parts $\kappa_0\ge\kappa_1\ge\dots\ge\kappa_z>0$. 
The Young subgroup $\Sigma_{[\kappa]}$ is a direct product of symmetric groups 
$\Sigma_{\kappa_l}$, $l=0,1,\dots,z$. Each group $\Sigma_{\kappa_l}$ contains permutations
of a $\kappa_l$-element set $\{K_l+1,K_l+2,\dots,K+\kappa_l\}$, 
where $K_l=\sum_{i=0}^{l-1}\kappa_i$. The nonordered partition $[0,0,2,2]$ is of a 
type represented by the ordered partition $[2,2]$ with a number of nonzero parts equal to 
$z+1=2$. Therefore, a class of conjugated orbits is represented by the Young subgroup
$\Sigma_2\otimes\Sigma_2$, where the first group $\Sigma_2$ contains permutations of 
the set $\{1,2\}$ ($K_0=0$) and the second factor $\Sigma_2$ --- of the set $\{3,4\}$
($K_1=2$). Other orbits of the same type are represented by states $\ket{a,a,b,b}$, 
where $a,b={-{3\over2}},{-{1\over2}},{1\over2},{3\over2}$ and $a\neq b$. A relation
between nonordered partitions $[k]$ of type $[2,2]$ and states representing orbits
$O[k]$ is presented below; the magnetization $M$ for states in a given orbit is
also calculated:
 \[
  \begin{array}{ccr}
  \mathrm{nonordered~partition} & \mathrm{orbit~representative} & M\\
\hline
    [2,2,0,0] & \ket{-3/2,-3/2,-1/2,-1/2} & -4 \\
  {}[2,0,2,0] & \ket{-3/2,-3/2,+1/2,+1/2} & -2 \\
  {}[2,0,0,2] & \ket{-3/2,-3/2,+3/2,+3/2} &  0 \\
  {}[0,2,2,0] & \ket{-1/2,-1/2,+1/2,+1/2} &  0 \\
  {}[0,2,0,2] & \ket{-1/2,-1/2,+3/2,+3/2} &  2 \\
  {}[0,0,2,2] & \ket{+1/2,+1/2,+3/2,+3/2} &  4 
\end{array} 
 \]

In the actual physical problems a symmetry group $G$ is, or at least can be embedded as, 
a subgroup of $\Sigma_n$. In a general case a restriction $\Sigma_n\downarrow G$ yields 
a decomposition of an orbit $O[k]$ into orbits of $G$. This decomposition depends on a 
type $[\kappa]$ of a nonordered partition, not on $[k]$ itself. In the example presented
each six-element orbit of type $[2,2]$ (determined by the action of $\Sigma_4$) 
is decomposed into two orbits of D$_4\subset\Sigma_4$. The first is represented by the 
state $\ket{a,a,b,b}$ and contains four states, since its stabilizer (in $\mathrm{D}_4$)
contains only the identity $E$ and the two-fold rotation $U_1$, which, as an element
of $\Sigma_4$, is a permutation $(12)(34)$;\cite{app2} since the stabilizer is two-element
subgroup of the eight-element group $\mathrm{D}_4$, then the orbit has $8/2=4$ elements:
$\ket{a,a,b,b}$, $\ket{a,b,b,a}$, $\ket{b,b,a,a}$, and $\ket{b,a,a,b}$. The other
two elements, $\ket{a,b,a,b}$ and $\ket{b,a,b,a}$, form a two-element orbit with the
stabilizer $\mathrm{D}_2^0=\{E,C_2,U_0,U_2\}$. All six orbits presented above have the
same decompositions into orbits of $\mathrm{D}_4$. More detailed discussion of combinatoric
properties of the system of four spin $s={3\over2}$ is presented in the Appendix.

Therefore, we have determined a few starting steps of our procedure: (i) generate
ordered partitions $[\kappa]$ of $n$ into no more than $2s+1$ nonzero parts; (ii) 
find a decomposition of an orbit $O[\kappa]$ into orbits of the Hamiltonian symmetry group $G$; 
(iii) for a chosen 
$-ns\le M\le ns$ determine all nonordered partitions $[k]$ satisfying the condition
(\ref{part}); (iv) decompositions of orbits $O[k]$ into orbits of
$G$ are analogous to those determined in Step~(ii).\cite{cpc01,app1} Note that  
orbits determined by the action of $G$ are collected into types labeled
by classes of conjugated subgroups in $G$ (in fact a representative $U\subseteq G$ is 
used as such a label; this subgroup is a stabilizer of an element
in an orbit under question). In this way an Ising state $\mu$ can be labeled by the following 
indices: magnetization $M$, a partition $[k]$, a stabilizer $U$, a representative $\nu$
of an orbit $G(\nu)\ni \mu$ ($G_\nu=U$), and a representative $g_r$ of a left coset 
$g_rU\subset G$ identifying $\mu$ in the orbit $G(\nu)$ ($g_r\nu=\mu$). These indices
play different roles in the further considerations: if a given Hamiltonian $\mathcal{H}$
commutes with $S^z$ then $M$ is a good quantum number and can be used as an additional
label of eigenspaces; $U$, $\nu$, and $g_r$ are used below to write down expressions for
matrix elements of spin operators (such as ${\mathcal H}$ or ${\mathbf S}^2$) in 
a compact and invariant form;\cite{cpc01,app1} a partition $[k]$ (together with an
ordered partition $[\kappa]$) is an additional index used during generation of Ising 
configurations and it is very useful in calculations of matrix elements.\cite{cpc01,cmstm} 

Now the linear structure of the space of states $L$ comes into play. The permutation 
representation $P$ restricted to an orbit of group $G$, and considered
as a vector (linear) representation of $G$, decomposes into irreps $\Gamma$ in the same
way for all orbits of a type $U$.\cite{app1} For example, there are two orbits leading
to $M=0$ and with D$_2\subset{\mathrm D}_4$ being a stabilizer represented by
$\ket{{3\over2},-{3\over2},{3\over2},-{3\over2}}$ and 
$\ket{{1\over2},-{1\over2},{1\over2},-{1\over2}}$, respectively  
($n=4$, $s={3\over2}$ as in the previous example). In both cases representation $P$ 
restricted to an orbit is decomposed into a direct sum $A_1\oplus B_1$. For each
orbit we construct the irreducible (symmetry adapted) basis 
  \[
    \ket{\Gamma v \gamma}=\sum_{r=1}^{|G|/|U|} a^{\Gamma v\gamma}_r \ket{g_r\nu}\,.
  \]
 Since at this moment the considerations are limited to one orbit $G(\nu)$, then
indices distinguishing orbits (the magnetization $M$ and the partition $[k]$) can be
omitted. The formula presented is valid for all orbits $G(\nu)$ of type $U$,
i.e.\ for orbits represented by a state $\nu$ with a stabilizer $G_\nu=U$.
 The index $v$ distinguishes copies of an irrep $\Gamma$ in the decomposition 
(\ref{decom}) restricted to an orbit $G(\nu)$. Note that $n(\Gamma)$ in such a case
is not larger than a dimension $[\Gamma]$ of an irrep $\Gamma$.\cite{ker,app1} The
coefficients $a^{\Gamma v\gamma}_r$ can be determined by the standard methods
used in the case of permutation representations.\cite{lullul} To obtain matrix elements
of any operator $H$ commuting with all $P(g)$ we have to combine all vectors labeled by 
a given irrep $\Gamma$ (including all copies). On the other hand, it follows from the 
Schur Lemma that it is enough to consider only one vector $\gamma$. It is rather tedious
than difficult to derive a general formula for matrix elements $H_{U\nu v,U'\nu' v'}$ labeled
by orbit stabilizers $U$ and $U'$, orbit representatives $\nu$ and $\nu'$, and indices
$v$ and $v'$ distinguishing copies of $\Gamma$.\cite{cpc01,app1} The formula obtained
contains products of two factors: a matrix element $\bra{U\nu g_r}H\ket{U'\nu'e_G}$ and 
a group-theoretical parameter of a model under discussion. Determination of the first
one is a numerical problem, so it is not discussed here. More details can be found in other
papers of authors.\cite{cpc01,cmstm} The second factor depends 
only on the symmetry group $G$ and can be determined once for all models with
a given symmetry group. Moreover, it is possible to determine such factors in an
analytical form for a family of groups, for example it has been done recently for
the dihedral groups ${\mathrm D}_n$ describing symmetry of molecular rings.\cite{app2} 

The ``magnetic'' step of the procedure proposed can be applied to models with
dominating Heisenberg term (\ref{ham}). At first we take into account the magnetization
$M$ restricting all considerations to subspaces $L_M$. We determine matrices of
the operator ${\mathbf S}^2$ for all irreps $\Gamma$ and solve eigenproblems 
${\mathbf S}^2\ket{\psi}=S(S+1)\ket{\psi}$ for $0\le S\le ns$. In fact, since each 
$S$-multiplet contains a state with $M=0$ (if $ns$ is integer), then states $\ket{SM}$ 
can be obtained from states $\ket{S\,0}$ acting with the total step operators 
$S^\pm=\sum_{j=1}^ns_j^\pm$. If we are interested in 
the ground state of a bipartite antiferromagnet only, then we can limit calculations to 
the case $S=0$ and the one-dimensional irrep. In most cases this is the unit  
representation $\Gamma_0$ ($A_1$ in the case of dihedral groups). However, if $s$ is 
half-integer and $n$ is not divisible by four, then the ground state is labeled by
the representation usually denoted as $\Gamma_1$ ($B_1$ for dihedral groups). 
This follows from the Marshall criterion:\cite{marsh,casp} Ising configurations related by 
the action of any pair of operators $s_j^+s_k^-$ have coefficients with opposite
signs in the decomposition of the ground state into the basic states.
In the case considered the ground state has to contain the N\'eel 
configuration $\ket{\tN1}=\ket{s,-s,\dots,-s}$. According with the Marshall rule the 
other N\'eel state, $\ket{\tN2}=\ket{{-s},s,\dots,s}$,
should enter the ground state with the same (opposite) sign if $ns$ is even (odd). 
Hence, the ground state has to contain the state 
$\ket{\tNe}=(\ket{\tN1}\pm\ket{\tN2})/\sqrt2$ 
which behaves as the basis vector of $\Gamma_0$ ($\Gamma_1$, respectively). 
 
\section{Eigenproblems and their solutions for the small ferric wheel}\label{fe6-eigen}

The dimension of the subspace $L_0$ in the case of six spins $s={5\over2}$ is equal
to 4332. Since $\dim L_1=4221$ then there are 111($=4332-4221$) states with 
$S=0$.\cite{app1,wald00} At first we determine such non-ordered partitions 
$[k]=[k_0,k_1,\dots,k_5]$ of $N=6$ into $2s+1=6$ parts that 
 \[ 
   \sum_{l=0}^5 k_l(l-{\textstyle{5\over2}}) = 0\quad \Leftrightarrow\quad
   \sum_{l=0}^5 k_l l = 15\,. 
 \]
  There are 11 ordered partitions of $N=6$ but only seven of them lead
to non-ordered partitions satisfying the condition presented above.
An algorithm discussed in Ref.~\onlinecite{cpc01} generates them in the 
following order: [1,1,1,1,1,1], [2,2,1,1], [3,1,1,1], [3,2,1], 
[4,1,1], [3,3], and [5,1]; these (ordered) partitions represent types of orbits 
of the symmetric group $\Sigma_6$. There are 1, 14, 6, 4, 2, 3, and 2 orbits 
(of states with $M=0$) labeled by non-ordered partitions represented by each of the seven 
ordered partitions, respectively (together 32 orbits). For example, the ordered
partition $[3,1,1,1]$ represents six nonordered partitions $[k_0,k_1,\dots,k_5]$
such that $\sum k_l l=15$; these partitions and representatives of the corresponding 
orbit $O[k]$ are as follows:
 \begin{eqnarray*}
  [1,0,1,3,1,0] &\colon&  \ket{ 
     {-\textstyle{5\over2}}, {-\textstyle{1\over2}}, { \textstyle{1\over2}},
     { \textstyle{1\over2}}, { \textstyle{1\over2}}, { \textstyle{3\over2}} }\,,\\
{}[1,1,1,0,3,0] &\colon&  \ket{ 
     {-\textstyle{5\over2}}, {-\textstyle{3\over2}}, {-\textstyle{1\over2}},
     { \textstyle{3\over2}}, { \textstyle{3\over2}}, { \textstyle{3\over2}} }\,,\\
{}[0,1,3,1,0,1] &\colon&  \ket{ 
     {-\textstyle{3\over2}}, {-\textstyle{1\over2}}, {-\textstyle{1\over2}},
     {-\textstyle{1\over2}}, { \textstyle{1\over2}}, { \textstyle{5\over2}} }\,,\\
{}[1,1,0,3,0,1] &\colon&  \ket{ 
     {-\textstyle{5\over2}}, {-\textstyle{3\over2}}, { \textstyle{1\over2}},
     { \textstyle{1\over2}}, { \textstyle{1\over2}}, { \textstyle{5\over2}} }\,,\\
{}[1,0,3,0,1,1] &\colon&  \ket{ 
     {-\textstyle{5\over2}}, {-\textstyle{1\over2}}, {-\textstyle{1\over2}},
     {-\textstyle{1\over2}}, { \textstyle{3\over2}}, { \textstyle{5\over2}} }\,,\\
{}[0,3,0,1,1,1] &\colon&  \ket{ 
     {-\textstyle{3\over2}}, {-\textstyle{3\over2}}, {-\textstyle{3\over2}},
     { \textstyle{1\over2}}, { \textstyle{3\over2}}, { \textstyle{5\over2}} }\,.
 \end{eqnarray*}
Taking into account orbit cardinalities, determined as 
$6!/k_0!k_1!k_2!k_3!k_4!k_5!$, one obtains
 \[ 1*720+14*180+6*120+4*60+2*30+3*20+2*6=4332 \]
 Ising configurations with $M=0$ what agrees with the number given above. 

We consider the nearest-neighbor interactions in a ring, so Eq.~(\ref{ham}) can 
be rewritten as
  \begin{equation}\label{hamlin}
    {\mathcal H}=J\sum_{j=1}^6 {\mathbf s}_j\cdot{\mathbf s}_{j+1}\,,\qquad
    j+6\equiv j\,.
  \end{equation}
The Hamiltonian point-symmetry group is $G={\mathrm S}_6\simeq {\mathrm D}_6$.\cite{fe06a} 
There are 10 classes of conjugated subgroups for this group, but only
seven of them can play a role of an orbit stabilizer.\cite{app2}
In the case considered, i.e.\ for the partitions presented above,
only three subgroups appear in the decompositions of orbits of $\Sigma_6$:
${\mathrm C}_1$, ${\mathrm D}_1^0$, and ${\mathrm D}_3^0$. These 
decompositions
are presented in Table~\ref{dcmp}. For example, each of six 120-element orbit of type 
$[3,1,1,1]$ decomposes into ten 12-element orbits with the trivial stabilizer $\mathrm{C}_1$.
Taking into account numbers of orbits
of $\Sigma_6$ one obtains:
339 orbits with a stabilizer ${\mathrm C}_1$, 43
orbits with a stabilizer ${\mathrm D}_1^0$, and 3
orbits with a stabilizer ${\mathrm D}_3^0$. These are all 385 orbits leading
to the total magnetization $M=0$. 

\begin{table}
\caption{Decompositions of orbits of the symmetric group $\Sigma_6$,
labeled by the partitions $[\kappa]$, into orbits of the dihedral group 
${\mathrm D}_6$, labeled by the stabilizers $U$. Each entry denotes 
number of orbits of type $U$ contained in an orbit of type $[\kappa]$. 
\label{dcmp}}

\begin{ruledtabular}
\begin{tabular}{c*{7}{r}}
~~~~$[\kappa]$ & [1,1,1,1,1,1] & [2,2,1,1] & [3,1,1,1] 
  & [3,2,1] & [4,1,1] & [3,3] & [5,1] \\[-4pt]
$U$~~~~ & & & & & & & \\ \hline
${\mathrm C}_1$~~~~   & 60 & 14 & 10 & 4 & 2 & 1 & 0 \\
${\mathrm D}_1^0$~~~~ &  0 &  2 &  0 & 2 & 1 & 1 & 1 \\
${\mathrm D}_3^0$~~~~ &  0 &  0 &  0 & 0 & 0 & 1 & 0 
\end{tabular}
\end{ruledtabular}
\end{table}

In the next step we take into account decompositions of
transitive representations $R^{{\mathrm D}_6:U}$ (i.e.\ the permutation representation
$P$ restricted to an orbit of type $U$) into irreps 
$\Gamma$ of ${\mathrm D}_6$.\cite{app2} The three 
stabilizers mentioned above give the following decompositions:
 \begin{eqnarray*}
  R^{{\mathrm D}_6:\mathrm{C_1}} &=& 
    A_1\oplus A_2 \oplus B_1 \oplus B_2 \oplus 2E_1 \oplus 2E_2\,;\\
  R^{{\mathrm D}_6:\mathrm{D_1}} &=& 
    A_1 \oplus B_1 \oplus E_1 \oplus E_2\,;\\
  R^{{\mathrm D}_6:\mathrm{D_3}} &=& A_1 \oplus B_1 \,.
 \end{eqnarray*}
This completes combinatorial and group-theoretical
classifications of states with $M=0$. In a similar way one can classify
states with $M=1$, what enables us to determine irreducible representations
related to $S=0$.\cite{app1} The numbers $n(\Gamma)$ of subspaces with a given
symmetry $\Gamma$ for $M=0,1$ and $S=0$ are collected in Table~\ref{ngamma}. 

\begin{table}
\caption{Numbers $n(\Gamma)$ for $M=0,1$ and $S=0$; the last one is simply
calculated as the difference of the previous two.
\label{ngamma}}

\begin{ruledtabular}
\begin{tabular}{c*{6}{r}}
$\Gamma$ & $A_1$ & $A_2$ & $B_1$ & $B_2$ & $E_1$ & $E_2$ \\ \hline
$n(\Gamma)$, $M=0$ & 385 & 339 & 385 & 339 & 721 & 721 \\
$n(\Gamma)$, $M=1$ & 383 & 325 & 365 & 334 & 699 & 708 \\ \hline
$n(\Gamma)$, $S=0$ &   2 &  14 &  20 &   5 &  22 &  13 
\end{tabular}
\end{ruledtabular}
\end{table}

Using the formulas discussed in Sec.~\ref{class}, and presented in the previous 
papers,\cite{app1,app2} matrices of the Hamiltonian ${\mathcal H}$ and the squared 
total spin ${\mathbf S}^2$ can be constructed. Of course, one can construct matrices for each 
$\Gamma=A_1,A_2,B_1,B_2,E_1,E_2$. However, considering the ground state 
it is enough to consider $\Gamma=B_1$: the ground state is a linear combination of 
385 Ising states with $M=0$ transforming as $\ket{B_1 b_1}$. We take into account also
$\Gamma=A_1$ since, starting from a theorem proved by Lieb and Mattis,\cite{lm,lsm} 
it can be shown that these representations label alternately states
with the lowest energy for $S=0,1,2,\dots,15$ ($A_1$ for odd and $B_1$ for even $S$,
respectively). The eigenstates determined for all $S$ and $\Gamma=A_1,B_1$ are used
to transform the Hamiltonian matrix to a block (quasi-diagonal) form. For example,
to find the ground state and its energy we have to solve 20-dimensional
eigenproblem (cf.\ Table~\ref{ngamma}).

A short note on numerical solutions are in place here. In the case of ${\mathbf S}^2$
operator and one-dimensional irreps eigenproblems can be solved {\it exactly}. It
has been achieved using the integer functions included in the {\sf GMP} package.\cite{gmp}
The square roots yielded by the action of $s_j^\pm$ operators and following
from the formulas for matrix elements,\cite{app2} can be easily removed
by the appropriate substitution.\cite{cpc01} In this way each eigenproblem for 
${\mathbf S}^2$ can be written as a system of homogeneous linear equations 
 with integer coefficients. Therefore, the solution(s) can be expressed by elements of 
the smallest number field containing integers, so---in a general case---by rational numbers
also easily maintained with the use of the {\sf GMP} package. 
The Hamiltonian matrix, with entries 
calculated with the {\sf GMP}-functions, is transformed to a block-form using the 
float functions from this package.\cite{gmp} The default precision is set 
to 256 bits. At this moment a procedure for solving eigenproblems using the float 
{\sf GMP}-functions has not been finished, so the eigenvalues and eigenvectors are 
determined using the standard procedures on 64-bit IRIX computers. The largest, 
60-dimensional, eigenproblem has been reached for $S=4$ and $\Gamma=B_1$.
Obtained vectors are verified checking their scalar products (including norms) and
their eigenvalues for ${\mathbf S}^2$ and ${\mathcal H}$ matrices. 

In this way
we have obtained 770 states expressed as linear combinations of the Ising configurations.
Since states with $S>0$ represent in fact multiplets then energies of 8436 states have been 
determined (4291 $A_1$ and 4145 $B_1$ states). The ground state energy is equal to
$E_0=-43.93471J$; for other values of the total spin $S$ the lowest energies
are in good agreement with Lande's rule $\Delta_S=(E_0(S)-E_0)/J=S(S+1)/3$, which approximates
the energy gaps.\cite{fe06a,fe06c} The largest relative deviation (3.75\%) 
is reached for $S=1$ when Lande's rule gives $\Delta_1={2\over3}$, whereas the value obtained
is equal to 0.69169. The other gaps ($S=2,3,\dots,15$) are 2.0744, 4.1469, 6.9070,
10.3517, 14.4772, 19.2792, 24.7534, 30.8952, 37.7002, 45.1633, 53.2789,
62.0396, 71.4347, 81.4347 and they agree with the numbers given in Ref.~\onlinecite{wald01}.
Though we have calculated about 18\% of eigenenergies we have obtained  
only a small fraction of low-lying states. For example, we have not determined energies
of the first excited state for each $S<15$. Hence, only a very rough approximation of 
the low-temperature
specific heat is calculated from the first four multiplets ($S\le 3$).
The best fit is 
achieved for $(E_0(1)-E_0)/k_{\text B}\approx 19.2$\,K, as in Ref.~\onlinecite{fe06b}.
Since $\Delta_1$ has been calculated to be about $0.69169$ we obtain
$J/k_{\text B}\approx 27.76$\,K (the value of 28.8\,K given in Ref.~\onlinecite{fe06a}
is obtained from Lande's rule, i.e.\ for $\Delta_1={2\over3}$). 
 
The quotient $E_0/J$ gives the total spin correlation
of the nearest neighbors. To compare it with other spin systems it
has to be divided by the maximum eigenvalue of a product 
$n{\mathbf s}_j\cdot{\mathbf s}_{j+1}$, 
i.e.\  by 52.5 for $s={5\over2}$, $n=6$. Hence we obtain 
$E_0/6Js(s+1)\approx -0.83685$, what means quite strong correlation.

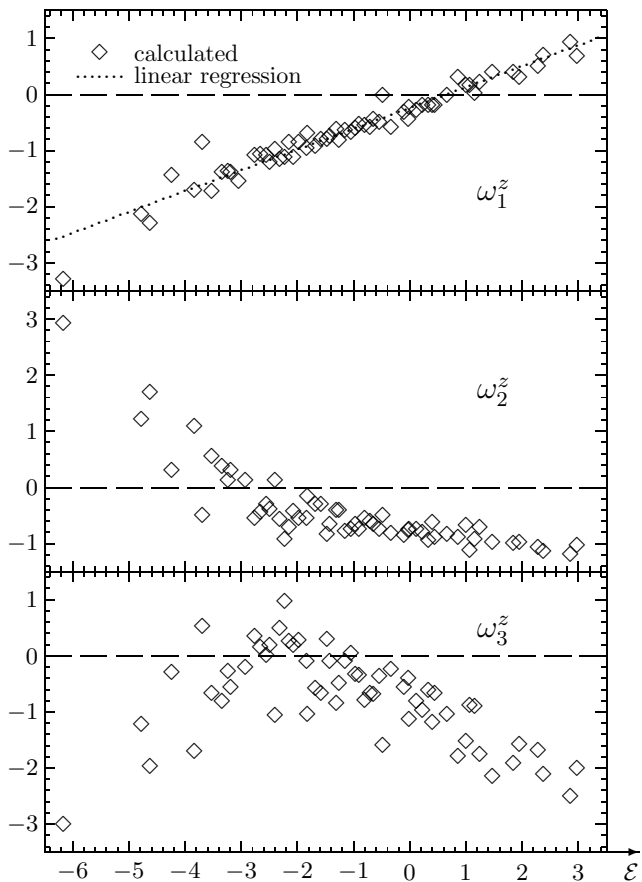
\begin{figure}[bth]
\setlength{\unitlength}{0.25pt}
\begin{picture}(960,1305)(0,60) 
\font\gnuplot=cmr10 at 9pt
\gnuplot
\thinlines
\linethickness {0.25pt}
\put(59.5,89.5){\vector(1,0){900}}
\multiput(59.5,514.5)(0,425){3}{\line(1,0){850}}
\multiput(59.5,89.5)(850,0){2}{\line(0,1){1275}}
\multiput(59.5,1237)(54,0){16}{\line(1,0){40}}
\multiput(59.5, 642)(54,0){16}{\line(1,0){40}}
\multiput(59.5, 387)(54,0){16}{\line(1,0){40}}

\multiput(60,  98)(0,17){2}{\line(1,0){ 7}}
\multiput(60,149)(0,85){14}{\multiput(0.0,0.0)(0,17){4}{\line(1,0){ 7}}}
\multiput(60,1339)(0,17){2}{\line(1,0){ 7}}
\multiput(60, 132)(0,85){15}{\line(1,0){15}}

\multiput(909,  98)(0,17){2}{\line(-1,0){ 7}}
\multiput(909, 149)(0,85){14}{\multiput(0.0,0.0)(0,17){4}{\line(-1,0){ 7}}}
\multiput(909,1339)(0,17){2}{\line(-1,0){ 7}}
\multiput(909, 132)(0,85){15}{\line(-1,0){15}}

\multiput(119, 90)(85,0){9}{\multiput(0.0,0.0)(17,0){4}{\line(0,1){ 7}}}
\multiput( 68, 90)(17,0){2}{\line(0,1){ 7}}
\multiput(884, 90)(17,0){2}{\line(0,1){ 7}}
\multiput(102, 90)(85,0){10}{\line(0,1){15}}

\multiput(119,1364)(85,0){9}{\multiput(0.0,0.0)(17,0){4}{\line(0,-1){ 7}}}
\multiput( 68,1364)(17,0){2}{\line(0,-1){ 7}}
\multiput(884,1364)(17,0){2}{\line(0,-1){ 7}}
\multiput(102,1364)(85,0){10}{\line(0,-1){15}}

\multiput(119,508)(85,0){9}{\multiput(0.0,0.0)(17,0){4}{\line(0,1){14}}}
\multiput( 68,508)(17,0){2}{\line(0,1){14}}
\multiput(884,508)(17,0){2}{\line(0,1){14}}
\multiput(102,500)(85,0){10}{\line(0,1){30}}

\multiput(119,933)(85,0){9}{\multiput(0.0,0.0)(17,0){4}{\line(0,1){14}}}
\multiput( 68,933)(17,0){2}{\line(0,1){14}}
\multiput(884,933)(17,0){2}{\line(0,1){14}}
\multiput(102,925)(85,0){10}{\line(0,1){30}}

\put(102,60){\makebox(0,0){$-6$}}
\put(187,60){\makebox(0,0){$-5$}}
\put(272,60){\makebox(0,0){$-4$}}
\put(357,60){\makebox(0,0){$-3$}}
\put(442,60){\makebox(0,0){$-2$}}
\put(527,60){\makebox(0,0){$-1$}}
\put(612,60){\makebox(0,0){$0$}}
\put(697,60){\makebox(0,0){$1$}}
\put(782,60){\makebox(0,0){$2$}}
\put(867,60){\makebox(0,0){$3$}}
\put(959,60){\makebox(0,0)[r]{${\mathcal E}$}}


\put(50, 982){\makebox(0,0)[r]{$-3$}}
\put(50,1067){\makebox(0,0)[r]{$-2$}}
\put(50,1152){\makebox(0,0)[r]{$-1$}}
\put(50,1237){\makebox(0,0)[r]{$0$}}
\put(50,1322){\makebox(0,0)[r]{$1$}}

\put(50, 557){\makebox(0,0)[r]{$-1$}}
\put(50, 642){\makebox(0,0)[r]{$0$}}
\put(50, 727){\makebox(0,0)[r]{$1$}}
\put(50, 812){\makebox(0,0)[r]{$2$}}
\put(50, 897){\makebox(0,0)[r]{$3$}}

\put(50, 132){\makebox(0,0)[r]{$-3$}}
\put(50, 217){\makebox(0,0)[r]{$-2$}}
\put(50, 302){\makebox(0,0)[r]{$-1$}}
\put(50, 387){\makebox(0,0)[r]{$0$}}
\put(50, 472){\makebox(0,0)[r]{$1$}}

\put(739,1087){\makebox(0,0){\large $\omega^z_1$}}
\put(739, 787){\makebox(0,0){\large $\omega^z_2$}}
\put(739, 404){\makebox(0,0)[b]{\large $\omega^z_3$}}

\put(143,1301){\raisebox{-.8pt}{\makebox(0,0){$\Diamond$}}}
\multiput(110,1264)(11,0){7}{\makebox(0,0){.}}
\put(195,1301){\makebox(0,0)[l]{calculated}}
\put(195,1264){\makebox(0,0)[l]{linear regression}}
 
\put( 88, 957){\raisebox{-.8pt}{\makebox(0,0){$\Diamond$}}}
\put(206,1056){\raisebox{-.8pt}{\makebox(0,0){$\Diamond$}}}
\put(219,1042){\raisebox{-.8pt}{\makebox(0,0){$\Diamond$}}}
\put(252,1114){\raisebox{-.8pt}{\makebox(0,0){$\Diamond$}}}
\put(286,1091){\raisebox{-.8pt}{\makebox(0,0){$\Diamond$}}}
\put(298,1165){\raisebox{-.8pt}{\makebox(0,0){$\Diamond$}}}
\put(312,1090){\raisebox{-.8pt}{\makebox(0,0){$\Diamond$}}}
\put(328,1119){\raisebox{-.8pt}{\makebox(0,0){$\Diamond$}}}
\put(337,1120){\raisebox{-.8pt}{\makebox(0,0){$\Diamond$}}}
\put(341,1119){\raisebox{-.8pt}{\makebox(0,0){$\Diamond$}}}
\put(353,1105){\raisebox{-.8pt}{\makebox(0,0){$\Diamond$}}}
\put(377,1145){\raisebox{-.8pt}{\makebox(0,0){$\Diamond$}}}
\put(386,1147){\raisebox{-.8pt}{\makebox(0,0){$\Diamond$}}}
\put(395,1144){\raisebox{-.8pt}{\makebox(0,0){$\Diamond$}}}
\put(400,1135){\raisebox{-.8pt}{\makebox(0,0){$\Diamond$}}}
\put(408,1154){\raisebox{-.8pt}{\makebox(0,0){$\Diamond$}}}
\put(415,1139){\raisebox{-.8pt}{\makebox(0,0){$\Diamond$}}}
\put(423,1142){\raisebox{-.8pt}{\makebox(0,0){$\Diamond$}}}
\put(429,1165){\raisebox{-.8pt}{\makebox(0,0){$\Diamond$}}}
\put(814,1296){\raisebox{-.8pt}{\makebox(0,0){$\Diamond$}}}
\put(436,1141){\raisebox{-.8pt}{\makebox(0,0){$\Diamond$}}}
\put(444,1165){\raisebox{-.8pt}{\makebox(0,0){$\Diamond$}}}
\put(456,1156){\raisebox{-.8pt}{\makebox(0,0){$\Diamond$}}}
\put(457,1178){\raisebox{-.8pt}{\makebox(0,0){$\Diamond$}}}
\put(469,1158){\raisebox{-.8pt}{\makebox(0,0){$\Diamond$}}}
\put(478,1169){\raisebox{-.8pt}{\makebox(0,0){$\Diamond$}}}
\put(487,1169){\raisebox{-.8pt}{\makebox(0,0){$\Diamond$}}}
\put(491,1174){\raisebox{-.8pt}{\makebox(0,0){$\Diamond$}}}
\put(501,1184){\raisebox{-.8pt}{\makebox(0,0){$\Diamond$}}}
\put(505,1167){\raisebox{-.8pt}{\makebox(0,0){$\Diamond$}}}
\put(514,1183){\raisebox{-.8pt}{\makebox(0,0){$\Diamond$}}}
\put(523,1179){\raisebox{-.8pt}{\makebox(0,0){$\Diamond$}}}
\put(530,1185){\raisebox{-.8pt}{\makebox(0,0){$\Diamond$}}}
\put(535,1192){\raisebox{-.8pt}{\makebox(0,0){$\Diamond$}}}
\put(544,1190){\raisebox{-.8pt}{\makebox(0,0){$\Diamond$}}}
\put(552,1187){\raisebox{-.8pt}{\makebox(0,0){$\Diamond$}}}
\put(557,1200){\raisebox{-.8pt}{\makebox(0,0){$\Diamond$}}}
\put(566,1195){\raisebox{-.8pt}{\makebox(0,0){$\Diamond$}}}
\put(571,1235){\raisebox{-.8pt}{\makebox(0,0){$\Diamond$}}}
\put(584,1188){\raisebox{-.8pt}{\makebox(0,0){$\Diamond$}}}
\put(603,1210){\raisebox{-.8pt}{\makebox(0,0){$\Diamond$}}}
\put(610,1200){\raisebox{-.8pt}{\makebox(0,0){$\Diamond$}}}
\put(611,1218){\raisebox{-.8pt}{\makebox(0,0){$\Diamond$}}}
\put(622,1213){\raisebox{-.8pt}{\makebox(0,0){$\Diamond$}}}
\put(631,1220){\raisebox{-.8pt}{\makebox(0,0){$\Diamond$}}}
\put(640,1220){\raisebox{-.8pt}{\makebox(0,0){$\Diamond$}}}
\put(646,1221){\raisebox{-.8pt}{\makebox(0,0){$\Diamond$}}}
\put(650,1220){\raisebox{-.8pt}{\makebox(0,0){$\Diamond$}}}
\put(668,1235){\raisebox{-.8pt}{\makebox(0,0){$\Diamond$}}}
\put(685,1263){\raisebox{-.8pt}{\makebox(0,0){$\Diamond$}}}
\put(697,1250){\raisebox{-.8pt}{\makebox(0,0){$\Diamond$}}}
\put(703,1250){\raisebox{-.8pt}{\makebox(0,0){$\Diamond$}}}
\put(710,1239){\raisebox{-.8pt}{\makebox(0,0){$\Diamond$}}}
\put(718,1255){\raisebox{-.8pt}{\makebox(0,0){$\Diamond$}}}
\put(737,1271){\raisebox{-.8pt}{\makebox(0,0){$\Diamond$}}}
\put(769,1271){\raisebox{-.8pt}{\makebox(0,0){$\Diamond$}}}
\put(778,1263){\raisebox{-.8pt}{\makebox(0,0){$\Diamond$}}}
\put(806,1280){\raisebox{-.8pt}{\makebox(0,0){$\Diamond$}}}
\put(855,1316){\raisebox{-.8pt}{\makebox(0,0){$\Diamond$}}}
\put(865,1295){\raisebox{-.8pt}{\makebox(0,0){$\Diamond$}}}

\multiput( 77.206, 1018.58)(10.794,4.000){77}{\makebox(0,0){.}}

\put( 88, 890){\raisebox{-.8pt}{\makebox(0,0){$\Diamond$}}}
\put(206, 745){\raisebox{-.8pt}{\makebox(0,0){$\Diamond$}}}
\put(219, 786){\raisebox{-.8pt}{\makebox(0,0){$\Diamond$}}}
\put(252, 667){\raisebox{-.8pt}{\makebox(0,0){$\Diamond$}}}
\put(286, 734){\raisebox{-.8pt}{\makebox(0,0){$\Diamond$}}}
\put(298, 600){\raisebox{-.8pt}{\makebox(0,0){$\Diamond$}}}
\put(312, 689){\raisebox{-.8pt}{\makebox(0,0){$\Diamond$}}}
\put(328, 674){\raisebox{-.8pt}{\makebox(0,0){$\Diamond$}}}
\put(337, 652){\raisebox{-.8pt}{\makebox(0,0){$\Diamond$}}}
\put(341, 667){\raisebox{-.8pt}{\makebox(0,0){$\Diamond$}}}
\put(363, 653){\raisebox{-.8pt}{\makebox(0,0){$\Diamond$}}}
\put(865, 554){\raisebox{-.8pt}{\makebox(0,0){$\Diamond$}}}
\put(377, 595){\raisebox{-.8pt}{\makebox(0,0){$\Diamond$}}}
\put(855, 540){\raisebox{-.8pt}{\makebox(0,0){$\Diamond$}}}
\put(386, 604){\raisebox{-.8pt}{\makebox(0,0){$\Diamond$}}}
\put(395, 617){\raisebox{-.8pt}{\makebox(0,0){$\Diamond$}}}
\put(400, 609){\raisebox{-.8pt}{\makebox(0,0){$\Diamond$}}}
\put(408, 652){\raisebox{-.8pt}{\makebox(0,0){$\Diamond$}}}
\put(415, 593){\raisebox{-.8pt}{\makebox(0,0){$\Diamond$}}}
\put(423, 564){\raisebox{-.8pt}{\makebox(0,0){$\Diamond$}}}
\put(429, 582){\raisebox{-.8pt}{\makebox(0,0){$\Diamond$}}}
\put(814, 545){\raisebox{-.8pt}{\makebox(0,0){$\Diamond$}}}
\put(436, 606){\raisebox{-.8pt}{\makebox(0,0){$\Diamond$}}}
\put(806, 552){\raisebox{-.8pt}{\makebox(0,0){$\Diamond$}}}
\put(444, 595){\raisebox{-.8pt}{\makebox(0,0){$\Diamond$}}}
\put(456, 596){\raisebox{-.8pt}{\makebox(0,0){$\Diamond$}}}
\put(457, 628){\raisebox{-.8pt}{\makebox(0,0){$\Diamond$}}}
\put(778, 559){\raisebox{-.8pt}{\makebox(0,0){$\Diamond$}}}
\put(469, 617){\raisebox{-.8pt}{\makebox(0,0){$\Diamond$}}}
\put(769, 557){\raisebox{-.8pt}{\makebox(0,0){$\Diamond$}}}
\put(478, 616){\raisebox{-.8pt}{\makebox(0,0){$\Diamond$}}}
\put(487, 571){\raisebox{-.8pt}{\makebox(0,0){$\Diamond$}}}
\put(491, 586){\raisebox{-.8pt}{\makebox(0,0){$\Diamond$}}}
\put(501, 608){\raisebox{-.8pt}{\makebox(0,0){$\Diamond$}}}
\put(505, 607){\raisebox{-.8pt}{\makebox(0,0){$\Diamond$}}}
\put(737, 559){\raisebox{-.8pt}{\makebox(0,0){$\Diamond$}}}
\put(514, 576){\raisebox{-.8pt}{\makebox(0,0){$\Diamond$}}}
\put(523, 578){\raisebox{-.8pt}{\makebox(0,0){$\Diamond$}}}
\put(718, 581){\raisebox{-.8pt}{\makebox(0,0){$\Diamond$}}}
\put(530, 586){\raisebox{-.8pt}{\makebox(0,0){$\Diamond$}}}
\put(710, 563){\raisebox{-.8pt}{\makebox(0,0){$\Diamond$}}}
\put(535, 578){\raisebox{-.8pt}{\makebox(0,0){$\Diamond$}}}
\put(703, 547){\raisebox{-.8pt}{\makebox(0,0){$\Diamond$}}}
\put(544, 595){\raisebox{-.8pt}{\makebox(0,0){$\Diamond$}}}
\put(697, 585){\raisebox{-.8pt}{\makebox(0,0){$\Diamond$}}}
\put(552, 591){\raisebox{-.8pt}{\makebox(0,0){$\Diamond$}}}
\put(557, 586){\raisebox{-.8pt}{\makebox(0,0){$\Diamond$}}}
\put(685, 566){\raisebox{-.8pt}{\makebox(0,0){$\Diamond$}}}
\put(566, 578){\raisebox{-.8pt}{\makebox(0,0){$\Diamond$}}}
\put(571, 599){\raisebox{-.8pt}{\makebox(0,0){$\Diamond$}}}
\put(668, 570){\raisebox{-.8pt}{\makebox(0,0){$\Diamond$}}}
\put(584, 573){\raisebox{-.8pt}{\makebox(0,0){$\Diamond$}}}
\put(650, 567){\raisebox{-.8pt}{\makebox(0,0){$\Diamond$}}}
\put(646, 589){\raisebox{-.8pt}{\makebox(0,0){$\Diamond$}}}
\put(603, 569){\raisebox{-.8pt}{\makebox(0,0){$\Diamond$}}}
\put(640, 562){\raisebox{-.8pt}{\makebox(0,0){$\Diamond$}}}
\put(610, 577){\raisebox{-.8pt}{\makebox(0,0){$\Diamond$}}}
\put(611, 579){\raisebox{-.8pt}{\makebox(0,0){$\Diamond$}}}
\put(631, 574){\raisebox{-.8pt}{\makebox(0,0){$\Diamond$}}}
\put(622, 579){\raisebox{-.8pt}{\makebox(0,0){$\Diamond$}}}

\put( 88,132){\raisebox{-.8pt}{\makebox(0,0){$\Diamond$}}}
\put(206,283){\raisebox{-.8pt}{\makebox(0,0){$\Diamond$}}}
\put(219,219){\raisebox{-.8pt}{\makebox(0,0){$\Diamond$}}}
\put(252,362){\raisebox{-.8pt}{\makebox(0,0){$\Diamond$}}}
\put(286,242){\raisebox{-.8pt}{\makebox(0,0){$\Diamond$}}}
\put(298,432){\raisebox{-.8pt}{\makebox(0,0){$\Diamond$}}}
\put(312,330){\raisebox{-.8pt}{\makebox(0,0){$\Diamond$}}}
\put(328,318){\raisebox{-.8pt}{\makebox(0,0){$\Diamond$}}}
\put(337,363){\raisebox{-.8pt}{\makebox(0,0){$\Diamond$}}}
\put(341,340){\raisebox{-.8pt}{\makebox(0,0){$\Diamond$}}}
\put(363,369){\raisebox{-.8pt}{\makebox(0,0){$\Diamond$}}}
\put(865,217){\raisebox{-.8pt}{\makebox(0,0){$\Diamond$}}}
\put(377,416){\raisebox{-.8pt}{\makebox(0,0){$\Diamond$}}}
\put(855,174){\raisebox{-.8pt}{\makebox(0,0){$\Diamond$}}}
\put(386,400){\raisebox{-.8pt}{\makebox(0,0){$\Diamond$}}}
\put(395,387){\raisebox{-.8pt}{\makebox(0,0){$\Diamond$}}}
\put(400,402){\raisebox{-.8pt}{\makebox(0,0){$\Diamond$}}}
\put(408,296){\raisebox{-.8pt}{\makebox(0,0){$\Diamond$}}}
\put(415,428){\raisebox{-.8pt}{\makebox(0,0){$\Diamond$}}}
\put(423,470){\raisebox{-.8pt}{\makebox(0,0){$\Diamond$}}}
\put(429,409){\raisebox{-.8pt}{\makebox(0,0){$\Diamond$}}}
\put(814,207){\raisebox{-.8pt}{\makebox(0,0){$\Diamond$}}}
\put(436,403){\raisebox{-.8pt}{\makebox(0,0){$\Diamond$}}}
\put(806,243){\raisebox{-.8pt}{\makebox(0,0){$\Diamond$}}}
\put(444,410){\raisebox{-.8pt}{\makebox(0,0){$\Diamond$}}}
\put(456,378){\raisebox{-.8pt}{\makebox(0,0){$\Diamond$}}}
\put(457,299){\raisebox{-.8pt}{\makebox(0,0){$\Diamond$}}}
\put(778,252){\raisebox{-.8pt}{\makebox(0,0){$\Diamond$}}}
\put(469,337){\raisebox{-.8pt}{\makebox(0,0){$\Diamond$}}}
\put(769,224){\raisebox{-.8pt}{\makebox(0,0){$\Diamond$}}}
\put(478,330){\raisebox{-.8pt}{\makebox(0,0){$\Diamond$}}}
\put(487,411){\raisebox{-.8pt}{\makebox(0,0){$\Diamond$}}}
\put(491,379){\raisebox{-.8pt}{\makebox(0,0){$\Diamond$}}}
\put(501,315){\raisebox{-.8pt}{\makebox(0,0){$\Diamond$}}}
\put(505,345){\raisebox{-.8pt}{\makebox(0,0){$\Diamond$}}}
\put(737,204){\raisebox{-.8pt}{\makebox(0,0){$\Diamond$}}}
\put(514,378){\raisebox{-.8pt}{\makebox(0,0){$\Diamond$}}}
\put(523,390){\raisebox{-.8pt}{\makebox(0,0){$\Diamond$}}}
\put(718,238){\raisebox{-.8pt}{\makebox(0,0){$\Diamond$}}}
\put(530,358){\raisebox{-.8pt}{\makebox(0,0){$\Diamond$}}}
\put(710,311){\raisebox{-.8pt}{\makebox(0,0){$\Diamond$}}}
\put(535,357){\raisebox{-.8pt}{\makebox(0,0){$\Diamond$}}}
\put(703,312){\raisebox{-.8pt}{\makebox(0,0){$\Diamond$}}}
\put(544,319){\raisebox{-.8pt}{\makebox(0,0){$\Diamond$}}}
\put(697,257){\raisebox{-.8pt}{\makebox(0,0){$\Diamond$}}}
\put(552,331){\raisebox{-.8pt}{\makebox(0,0){$\Diamond$}}}
\put(557,329){\raisebox{-.8pt}{\makebox(0,0){$\Diamond$}}}
\put(685,235){\raisebox{-.8pt}{\makebox(0,0){$\Diamond$}}}
\put(566,356){\raisebox{-.8pt}{\makebox(0,0){$\Diamond$}}}
\put(571,251){\raisebox{-.8pt}{\makebox(0,0){$\Diamond$}}}
\put(668,298){\raisebox{-.8pt}{\makebox(0,0){$\Diamond$}}}
\put(584,367){\raisebox{-.8pt}{\makebox(0,0){$\Diamond$}}}
\put(650,330){\raisebox{-.8pt}{\makebox(0,0){$\Diamond$}}}
\put(646,287){\raisebox{-.8pt}{\makebox(0,0){$\Diamond$}}}
\put(603,339){\raisebox{-.8pt}{\makebox(0,0){$\Diamond$}}}
\put(640,334){\raisebox{-.8pt}{\makebox(0,0){$\Diamond$}}}
\put(610,353){\raisebox{-.8pt}{\makebox(0,0){$\Diamond$}}}
\put(611,291){\raisebox{-.8pt}{\makebox(0,0){$\Diamond$}}}
\put(631,304){\raisebox{-.8pt}{\makebox(0,0){$\Diamond$}}}
\put(622,318){\raisebox{-.8pt}{\makebox(0,0){$\Diamond$}}}

\end{picture}
\caption{Correlations vs energy per spin for $S=4,\Gamma=B_1$.\label{c04b1}}
\end{figure}

\section{Spin correlations and the classical ordering}\label{fe6-res}

The thermodynamic quantities have been already determined for the Fe$_6$ cluster
and the results have been compared with the experimental data.\cite{fe06a,fe06b,wald01} 
Therefore, in this paper we focus our attention on the eigenstates properties: 
the spin correlations $\omega^z_r=\langle\sum_{j=1}^6 s_j^zs_{j+r}^z\rangle/6$
for $r=1,2,3$ and the probability $p_{\tN}$ of finding the system in the N\'eel state.
The latter one is, of course, equal to $a_{\tN1}^2=a_{\tN2}^2$, i.e.\ to the square
of the coefficient of $\ket{\tN1}$ or $\ket{\tN2}$ in a state under consideration. It
has to be stressed that $a_{\tN1}$ is nonzero if $S$ is even (odd) and $\Gamma=A_1$
($B_1$, respectively). The value of $p_{\tN}$ can be used to compare quantum states
determined with the classical ordering of antiferromagnets.

\subsection{Spin correlations}\label{corr}

In the case $S=0$, due to isotropy in spin-space, the $z$-correlations of the nearest 
neighbors, $\omega^z_1$, should be one third of energy (per spin) in a given state.
This has been confirmed by the calculations presented elsewhere.\cite{flonewx} For
other states with $M=0$ ($\Gamma=A_1,B_1$, $0\le S\le 15$) we have plotted 
$\omega^z_r$ versus energy per spin ${\mathcal E}=E/6J$. Typical drawings are presented in 
Fig.~\ref{c04b1} for $S=4$ and $\Gamma=B_1$. In all cases $\omega^z_1$ is almost
linear function of ${\mathcal E}$ (with a positive slope) and only a few values for 
high energies are larger than zero. On the other hand, $\omega^z_2$ decreases with 
growing ${\mathcal E}$, but only some low-lying states show positive (ferromagnetic)
$\omega^z_2$ with negative $\omega^z_1$ and $\omega^z_3$, the feature characteristic 
for the ground state of an antiferromagnetic system. However, it might happen that
for $\omega^z_2<0$ spin correlations of the next-nearest neighbors are positive for other 
quantization axes. It could be checked considering terms 
$\sum_j{\mathbf s}_j\cdot{\mathbf s}_{j+2}$, what has not been done in our calculations.
We tried to fit some simple functions to $\omega^z_2({\mathcal E})$, but
the results obtained occurred rather ambiguous. It seems that an exponential function
$\omega^z_2({\mathcal E})+1.132=\exp({-1.224}-0.432{\mathcal E})$ fits better than the power 
law $\omega^z_2({\mathcal E})+3.191=6.850\cdot({\mathcal E}-{\mathcal E}_0)^{-0.532}$
($S=4$, $\Gamma=B_1$). The correlations with the third neighbors 
$\omega^z_3({\mathcal E})$ show 
positive maximum for all $S<14$, $\Gamma=A_1$ and $S<13$, $\Gamma=B_1$. It is interesting  
that single states, obtained for pairs $(S,\Gamma)=(13,B_1),(14,B_1), (15,A_1)$,
have $\omega^z_3$ negative: ${-0.1833},{-0.6327},{-0.2155}$, respectively; there are no 
states for pairs $(14,A_1)$ and $(15,B_1)$.

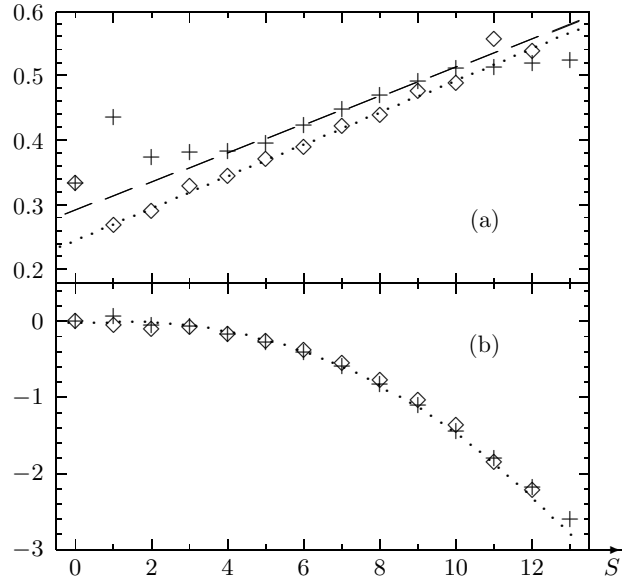
\begin{figure}[bt]
\setlength{\unitlength}{0.24pt}
\begin{picture}(960,888)(0,0)
\font\gnuplot=cmr10 at 9pt
\gnuplot
\linethickness{0.25pt}
\multiput( 70.0,482.0)(0,102){4}{\line(1,0){16}}
\multiput( 70.0,502.4)(0,102){4}{\multiput(0,0)(0,20.4){4}{\line(1,0){8}}}
\multiput(910.0,502.4)(0,102){4}{\multiput(0,0)(0,20.4){4}{\line(-1,0){8}}}
\multiput(910.0,482.0)(0,102){4}{\line(-1,0){16}}

\multiput(100,444)(120,0){7}{\line(0,1){32}}
\multiput(160,452)(120,0){7}{\line(0,1){16}}
\multiput(100,888)(120,0){7}{\line(0,-1){8}}
\multiput(160,888)(120,0){7}{\line(0,-1){16}}
\multiput(100, 40)(120,0){7}{\line(0,1){16}}
\multiput(160, 40)(120,0){7}{\line(0,1){8}}

\multiput( 70.0,160)(0,120){3}{\line(1,0){16}}
\multiput( 70.0,64)(0,120){3}{\multiput(0,0)(0,24){4}{\line(1,0){8}}}
\multiput( 70.0,424)(0,24){2}{\line(1,0){8}}
\multiput(910.0,64)(0,120){3}{\multiput(0,0)(0,24){4}{\line(-1,0){8}}}
\multiput(910.0,424)(0,24){2}{\line(-1,0){8}}
\multiput(910.0,160)(0,120){3}{\line(-1,0){16}}

\put( 70, 40){\line(0,1){848}}
\put(910, 40){\line(0,1){848}}
\put( 70,888){\line(1,0){840}}
\put( 70,460){\line(1,0){840}}

\put(50,480){\makebox(0,0)[r]{0.2}}
\put(50,582){\makebox(0,0)[r]{0.3}}
\put(50,684){\makebox(0,0)[r]{0.4}}
\put(50,786){\makebox(0,0)[r]{0.5}}
\put(50,888){\makebox(0,0)[r]{0.6}}
 
\put(50, 40){\makebox(0,0)[r]{$-3$}}
\put(50,160){\makebox(0,0)[r]{$-2$}}
\put(50,280){\makebox(0,0)[r]{$-1$}}
\put(50,400){\makebox(0,0)[r]{$0$}}

\put(100,0){\makebox(0,0)[b]{$0$}}
\put(220,0){\makebox(0,0)[b]{$2$}}
\put(340,0){\makebox(0,0)[b]{$4$}}
\put(460,0){\makebox(0,0)[b]{$6$}}
\put(580,0){\makebox(0,0)[b]{$8$}}
\put(700,0){\makebox(0,0)[b]{$10$}}
\put(820,0){\makebox(0,0)[b]{$12$}}
\put( 70,40){\vector(1,0){890}}
\put(959,0){\makebox(0,0)[br]{$S$}}

\put(770,540){\makebox(0,0)[br]{(a)}}
\put(770,380){\makebox(0,0)[tr]{(b)}}

\put(100,618){\raisebox{-.8pt}{\makebox(0,0){$\Diamond$}}}
\put(160,552){\raisebox{-.8pt}{\makebox(0,0){$\Diamond$}}}
\put(220,573){\raisebox{-.8pt}{\makebox(0,0){$\Diamond$}}}
\put(280,613){\raisebox{-.8pt}{\makebox(0,0){$\Diamond$}}}
\put(340,629){\raisebox{-.8pt}{\makebox(0,0){$\Diamond$}}}
\put(400,655){\raisebox{-.8pt}{\makebox(0,0){$\Diamond$}}}
\put(460,674){\raisebox{-.8pt}{\makebox(0,0){$\Diamond$}}}
\put(520,708){\raisebox{-.8pt}{\makebox(0,0){$\Diamond$}}}
\put(580,725){\raisebox{-.8pt}{\makebox(0,0){$\Diamond$}}}
\put(640,763){\raisebox{-.8pt}{\makebox(0,0){$\Diamond$}}}
\put(700,775){\raisebox{-.8pt}{\makebox(0,0){$\Diamond$}}}
\put(760,845){\raisebox{-.8pt}{\makebox(0,0){$\Diamond$}}}
\put(820,826){\raisebox{-.8pt}{\makebox(0,0){$\Diamond$}}}

\put(100,618){\makebox(0,0){$+$}}
\put(160,722){\makebox(0,0){$+$}}
\put(220,660){\makebox(0,0){$+$}}
\put(280,667){\makebox(0,0){$+$}}
\put(340,669){\makebox(0,0){$+$}}
\put(400,682){\makebox(0,0){$+$}}
\put(460,710){\makebox(0,0){$+$}}
\put(520,735){\makebox(0,0){$+$}}
\put(580,758){\makebox(0,0){$+$}}
\put(640,780){\makebox(0,0){$+$}}
\put(700,799){\makebox(0,0){$+$}}
\put(760,802){\makebox(0,0){$+$}}
\put(820,808){\makebox(0,0){$+$}}
\put(880,812){\makebox(0,0){$+$}}

\multiput( 76.333,517.84)(16.667,7){50}{\makebox(0,0){.}}
\multiput( 84.0,569.0)(64,24){13}{\line(5,2){42}}

\put(100,400){\makebox(0,0){$+$}}
\put(160,408){\makebox(0,0){$+$}}
\put(220,394){\makebox(0,0){$+$}}
\put(280,393){\makebox(0,0){$+$}}
\put(340,380){\makebox(0,0){$+$}}
\put(400,368){\makebox(0,0){$+$}}
\put(460,352){\makebox(0,0){$+$}}
\put(520,330){\makebox(0,0){$+$}}
\put(580,301){\makebox(0,0){$+$}}
\put(640,268){\makebox(0,0){$+$}}
\put(700,227){\makebox(0,0){$+$}}
\put(760,185){\makebox(0,0){$+$}}
\put(820,139){\makebox(0,0){$+$}}
\put(880, 89){\makebox(0,0){$+$}}
\put(100,400){\raisebox{-0.8pt}{\makebox(0,0){$\Diamond$}}}
\put(160,393){\raisebox{-0.8pt}{\makebox(0,0){$\Diamond$}}}
\put(220,388){\raisebox{-0.8pt}{\makebox(0,0){$\Diamond$}}}
\put(280,391){\raisebox{-0.8pt}{\makebox(0,0){$\Diamond$}}}
\put(340,379){\raisebox{-0.8pt}{\makebox(0,0){$\Diamond$}}}
\put(400,369){\raisebox{-0.8pt}{\makebox(0,0){$\Diamond$}}}
\put(460,354){\raisebox{-0.8pt}{\makebox(0,0){$\Diamond$}}}
\put(520,334){\raisebox{-0.8pt}{\makebox(0,0){$\Diamond$}}}
\put(580,307){\raisebox{-0.8pt}{\makebox(0,0){$\Diamond$}}}
\put(640,275){\raisebox{-0.8pt}{\makebox(0,0){$\Diamond$}}}
\put(700,236){\raisebox{-0.8pt}{\makebox(0,0){$\Diamond$}}}
\put(760,178){\raisebox{-0.8pt}{\makebox(0,0){$\Diamond$}}}
\put(820,134){\raisebox{-0.8pt}{\makebox(0,0){$\Diamond$}}}

\put( 90.0,398.50){\makebox(0,0){.}}
\put(110.0,398.50){\makebox(0,0){.}}
\put(130.0,398.50){\makebox(0,0){.}}
\put(150.0,398.50){\makebox(0,0){.}}
\put(170.0,399.00){\makebox(0,0){.}}
\put(190.0,399.00){\makebox(0,0){.}}
\put(210.0,399.00){\makebox(0,0){.}}
\put(230.0,398.06){\makebox(0,0){.}}
\put(250.0,396.65){\makebox(0,0){.}}
\put(270.0,395.35){\makebox(0,0){.}}
\put(290.0,392.90){\makebox(0,0){.}}
\put(310.0,389.64){\makebox(0,0){.}}
\put(330.0,386.29){\makebox(0,0){.}}
\put(350.0,382.82){\makebox(0,0){.}}
\put(370.0,378.32){\makebox(0,0){.}}
\put(390.0,372.97){\makebox(0,0){.}}
\put(410.0,367.93){\makebox(0,0){.}}
\put(430.0,362.28){\makebox(0,0){.}}
\put(450.0,355.42){\makebox(0,0){.}}
\put(470.0,348.85){\makebox(0,0){.}}
\put(490.0,341.99){\makebox(0,0){.}}
\put(506.0,334.49){\makebox(0,0){.}}
\put(524.0,326.45){\makebox(0,0){.}}
\put(543.0,318.66){\makebox(0,0){.}}
\put(561.5,310.08){\makebox(0,0){.}}
\put(579.0,300.51){\makebox(0,0){.}}
\put(591.5,291.26){\makebox(0,0){.}}
\put(612.0,281.52){\makebox(0,0){.}}
\put(628.5,272.15){\makebox(0,0){.}}
\put(645.5,261.55){\makebox(0,0){.}}
\put(661.0,250.80){\makebox(0,0){.}}
\put(677.0,240.17){\makebox(0,0){.}}
\put(692.5,228.72){\makebox(0,0){.}}
\put(708.0,217.38){\makebox(0,0){.}}
\put(723.0,205.78){\makebox(0,0){.}}
\put(738.0,193.90){\makebox(0,0){.}}
\put(752.5,181.29){\makebox(0,0){.}}
\put(767.5,169.11){\makebox(0,0){.}}
\put(782.0,156.57){\makebox(0,0){.}}
\put(796.5,143.79){\makebox(0,0){.}}
\put(811.0,130.76){\makebox(0,0){.}}
\put(825.0,117.83){\makebox(0,0){.}}
\put(839.0,104.37){\makebox(0,0){.}}
\put(852.5, 90.66){\makebox(0,0){.}}
\put(867.0, 77.41){\makebox(0,0){.}}
\put(881.5, 63.14){\makebox(0,0){.}}

\end{picture}
\caption{(a) A slope $a(S)$ and (b) an intercept $b(S)$ as a function of the total
spin $S$; `$\Diamond$' and `$+$': results of linear regression $\omega^z_1({\mathcal E})
=a(S){\mathcal E}+b(S)$; lines give (a) linear regressions of $a_0(S)$ (dashed line), $a_1(S)$
(dotted line) and (b) a parabolic fit to $b(S)$ (see text for details).\label{abspin}}
\end{figure}

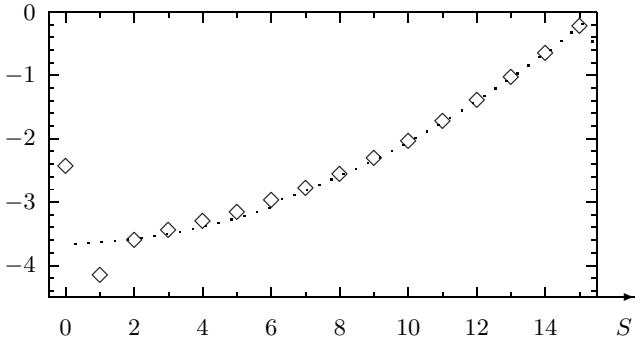
\begin{figure}
\setlength{\unitlength}{0.24pt}
\ifx\plotpoint\undefined\newsavebox{\plotpoint}\fi
\begin{picture}(994,510)(0,0)
\font\gnuplot=cmr10 at 9pt
\gnuplot
\sbox{\plotpoint}{\rule[-0.200pt]{0.600pt}{0.400pt}}%
\linethickness{0.25pt}

\put(50,110){\makebox(0,0)[r]{$-4$}}
\put(50,210){\makebox(0,0)[r]{$-3$}}
\put(50,310){\makebox(0,0)[r]{$-2$}}
\put(50,410){\makebox(0,0)[r]{$-1$}}
\put(50,510){\makebox(0,0)[r]{$0$}}
\put( 97, 0){\makebox(0,0)[b]{$ 0$}}
\put(205, 0){\makebox(0,0)[b]{$ 2$}}
\put(313, 0){\makebox(0,0)[b]{$ 4$}}
\put(421, 0){\makebox(0,0)[b]{$ 6$}}
\put(529, 0){\makebox(0,0)[b]{$ 8$}}
\put(637, 0){\makebox(0,0)[b]{$10$}}
\put(745, 0){\makebox(0,0)[b]{$12$}}
\put(853, 0){\makebox(0,0)[b]{$14$}}
\put(990, 0){\makebox(0,0)[br]{$S$}}

\multiput(70,110)(0,100){4}{\line(1,0){16}}
\multiput(934,110)(0,100){4}{\line(-1,0){16}}
\put(70, 70){\line(1,0){8}}
\put(70, 90){\line(1,0){8}}
\put(934, 70){\line(-1,0){8}}
\put(934, 90){\line(-1,0){8}}
\multiput(70,130)(0,100){4}{\multiput(0,0)(0,20){4}{\line(1,0){8}}}
\multiput(934,130)(0,100){4}{\multiput(0,0)(0,20){4}{\line(-1,0){8}}}

\put( 70, 60){\vector(1,0){924}}
\put( 70, 60){\line(0,1){450}}
\put(934, 60){\line(0,1){450}}
\put( 70,510){\line(1,0){864}}

\multiput( 97, 60)(108,0){8}{\line(0,1){16}}
\multiput(151, 60)(108,0){8}{\line(0,1){ 8}}
\multiput( 97,510)(108,0){8}{\line(0,-1){16}}
\multiput(151,510)(108,0){8}{\line(0,-1){ 8}}

\put( 97,266){\raisebox{-.8pt}{\makebox(0,0){$\Diamond$}}}
\put(151, 94){\raisebox{-.8pt}{\makebox(0,0){$\Diamond$}}}
\put(205,149){\raisebox{-.8pt}{\makebox(0,0){$\Diamond$}}}
\put(259,166){\raisebox{-.8pt}{\makebox(0,0){$\Diamond$}}}
\put(313,179){\raisebox{-.8pt}{\makebox(0,0){$\Diamond$}}}
\put(367,194){\raisebox{-.8pt}{\makebox(0,0){$\Diamond$}}}
\put(421,212){\raisebox{-.8pt}{\makebox(0,0){$\Diamond$}}}
\put(475,232){\raisebox{-.8pt}{\makebox(0,0){$\Diamond$}}}
\put(529,254){\raisebox{-.8pt}{\makebox(0,0){$\Diamond$}}}
\put(583,279){\raisebox{-.8pt}{\makebox(0,0){$\Diamond$}}}
\put(637,306){\raisebox{-.8pt}{\makebox(0,0){$\Diamond$}}}
\put(691,337){\raisebox{-.8pt}{\makebox(0,0){$\Diamond$}}}
\put(745,370){\raisebox{-.8pt}{\makebox(0,0){$\Diamond$}}}
\put(799,407){\raisebox{-.8pt}{\makebox(0,0){$\Diamond$}}}
\put(853,445){\raisebox{-.8pt}{\makebox(0,0){$\Diamond$}}}
\put(907,487){\raisebox{-.8pt}{\makebox(0,0){$\Diamond$}}}

\put(111.45,144.30){\usebox{\plotpoint}}
\put(132.14,145.50){\usebox{\plotpoint}}
\put(152.81,146.94){\usebox{\plotpoint}}
\put(173.48,148.56){\usebox{\plotpoint}}
\put(194.10,150.94){\usebox{\plotpoint}}
\put(214.73,153.39){\usebox{\plotpoint}}
\put(235.20,156.90){\usebox{\plotpoint}}
\put(255.75,159.88){\usebox{\plotpoint}}
\put(276.17,163.66){\usebox{\plotpoint}}
\put(296.54,167.71){\usebox{\plotpoint}}
\put(316.79,172.52){\usebox{\plotpoint}}
\put(337.02,177.49){\usebox{\plotpoint}}
\put(357.23,182.51){\usebox{\plotpoint}}
\put(377.23,188.33){\usebox{\plotpoint}}
\put(397.15,194.36){\usebox{\plotpoint}}
\put(416.90,201.07){\usebox{\plotpoint}}
\put(436.82,214.54){\usebox{\plotpoint}}
\put(456.40,221.50){\usebox{\plotpoint}}
\put(476.09,229.46){\usebox{\plotpoint}}
\put(495.43,236.77){\usebox{\plotpoint}}
\put(515.02,245.67){\usebox{\plotpoint}}
\put(534.18,254.53){\usebox{\plotpoint}}
\put(553.20,262.83){\usebox{\plotpoint}}
\put(572.31,271.91){\usebox{\plotpoint}}
\put(591.16,280.58){\usebox{\plotpoint}}
\put(610.07,291.14){\usebox{\plotpoint}}
\put(628.41,300.74){\usebox{\plotpoint}}
\put(646.97,310.99){\usebox{\plotpoint}}
\put(665.36,321.60){\usebox{\plotpoint}}
\put(683.78,332.11){\usebox{\plotpoint}}
\put(701.85,343.27){\usebox{\plotpoint}}
\put(720.17,354.98){\usebox{\plotpoint}}
\put(737.70,366.06){\usebox{\plotpoint}}
\put(755.73,377.33){\usebox{\plotpoint}}
\put(773.29,389.38){\usebox{\plotpoint}}
\put(790.69,401.61){\usebox{\plotpoint}}
\put(808.17,412.78){\usebox{\plotpoint}}
\put(825.59,425.56){\usebox{\plotpoint}}
\put(842.86,438.57){\usebox{\plotpoint}}
\put(859.81,450.54){\usebox{\plotpoint}}
\put(876.92,464.27){\usebox{\plotpoint}}
\put(893.54,477.69){\usebox{\plotpoint}}
\put(910.32,490.88){\usebox{\plotpoint}}
\put(926.71,463.55){\usebox{\plotpoint}}

\end{picture}
\caption{Correlations in the state with the lowest energy in each $S$-multiplet vs the
total spin $S$ (diamonds). The dotted line: Lande's rule for correlations 
given by Eq.~(\ref{Lrule}).\label{corrspin}}
\end{figure}

The correlation $\omega^z_1$ has been analyzed in two aspects. At first, assuming linear 
dependence $\omega^z_1({\mathcal E})=a(S){\mathcal E}+b(S)$ for each $S$, we have
plotted coefficients $a(S)$ and $b(S)$ versus $S$ (see Fig.~\ref{abspin}). 
Results of linear regressions are denoted by the crosses `$+$' are obtained for pairs 
$(S,\Gamma)$ containing the ground state of an $S$-multiplet, i.e.\ $S$ even, $\Gamma=B_1$ and 
$S$ odd, $\Gamma=A_1$; the diamonds `$\Diamond$' denote results for the other pairs.
Since $a(S)$ behaves differently in those cases, then the linear functions $a_0(S)$ 
(the first case, dashed line) and $a_1(S)$ (the other case, dotted line) have been
considered separately. The points for $S=0$ in both cases and for $S=1$ in the first case have 
been omitted due to highly irregular behavior (see also below and Fig.~\ref{corrspin}).
As the results we have obtained 
 \[
   a_0(S)=0.0221S+0.291\,,\quad
   a_1(S)=0.0249S+0.245\,.
 \]
 It has to be stressed that $a_0(S)$ deviates from a line for high $S$ and, maybe, a parabolic
fit should be applied in this region. The intercept $b(S)$ does not show such irregularities so 
all results are fitted to one (second order) function. The best-fit line is given as
 \[
   b(S)={-0.0197}S^2+0.0591S-0.0518\,.
 \]

Let $\omega_0(S)$ denote $\omega^z_1$ in the state with the lowest energy for each $S$-multiplet. 
The plot in Fig.~\ref{corrspin} shows this function fitted by a relation analogous to Lande's
rule
\begin{equation}\label{Lrule}
 \omega_0(S)={1\over72}S(S+1)+{1\over2}{\mathcal E}_0\,.
\end{equation}
Substituting Lande's rule $\Delta_S=S(S+1)/3$ one obtains
\begin{equation}\label{omen}
 \omega_0(S)={1\over4}({\mathcal E}_0(S)+{\mathcal E}_0)\,.
\end{equation}
 Lande's rule for correlations is broken for $S=0,1$. In the first case Eq.~(\ref{Lrule}) yields
$\omega_0(S)={1\over2}{\mathcal E}_0$, whereas the exact value is ${\mathcal E}_0/3$. The latter
case, $S=1$, shows deviation in the other direction: calculated value of $\omega_0(1)$
is ${-4.172}$, whereas from Eq.~(\ref{omen}) one obtains ${-3.632}$, so the real value
is about 15\% larger in magnitude. Similar results have been obtained for six spins ${1\over2}$
and ${3\over2}$.

\subsection{N\'eel probability}

It is interesting to compare the results obtained with correlations in the N\'eel
state. In this case $z$-correlations (per bond) are equal to $\pm 6.25$
since the N\'eel state exhibits the long-range order. The ground state determined here has
alternating, but decreasing in magnitude, correlations. However, the total nearest neighbors
correlation, equal to the ground state energy $3\omega_0(0)$, 
is larger than in the N\'eel 
state, since in the N\'eel state $x$- and $y$-correlations are equal to zero.  
The orbit containing the both
N\'eel states gives one state with symmetry $\Gamma=B_1$,
namely $|\tNe\rangle$. The absolute value of the coefficient 
of this state in the ground state is $a_N=0.3284$, so the probability is 
equal $p_{\tN}=0.1078$. The probability of a single N\'eel state is 
$p_{\tN1}=p_{\tN2}={1\over2}p_{\tN}=
0.0539$. Such small number does not explain so strong nearest neighbor
correlation in the ground state. However, one has to take into account that the N\'eel
state, $|\tN1\rangle$ or $|\tN2\rangle$, represents only one possibility of such spin arrangement
that each of sublattices (of a bipartite antiferromagnet) has the maximum total spin 
$S_{\text{max}}$ (in the other words---each sublattice is ordered ferromagnetically),
but magnetizations have opposite directions. For a given spin number, $S_{\text{max}}$ in 
the case considered, there are $2S_{\text{max}}+1$ possible spin projections, so there
are $2S_{\text{max}}+1$ different N\'eel states:\cite{casp} each of them
corresponds to different choice of quantization axis and its orientation. In the case 
considered $S_{\text{max}}=3\times{5\over2}={15\over2}$, then there are 16 N\'eel states.
Two of these states are given by $|\tN1\rangle$ and $|\tN2\rangle$, whereas the others can be 
obtained as linear combinations of states with $S=M=0$. Hence, the global probability is 
$p_{\tN}^*=16p_{\tN1}=0.8624$. 

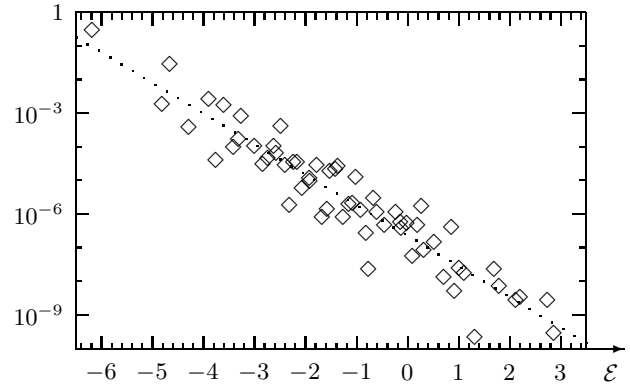
\begin{figure}
\setlength{\unitlength}{0.240900pt}
\ifx\plotpoint\undefined\newsavebox{\plotpoint}\fi
\begin{picture}(950,603)(52, 0)
\font\gnuplot=cmr10 at 9pt
\gnuplot
\sbox{\plotpoint}{\rule[-0.200pt]{0.600pt}{0.600pt}}%
\put(144.0,127.0){\rule[-0.200pt]{4.818pt}{0.400pt}}
\put(126,127){\makebox(0,0)[r]{$10^{-9}$}}
\put(925.0,127.0){\rule[-0.200pt]{4.818pt}{0.400pt}}
\put(144.0,180.0){\rule[-0.200pt]{2.409pt}{0.400pt}}
\put(935.0,180.0){\rule[-0.200pt]{2.409pt}{0.400pt}}
\put(144.0,233.0){\rule[-0.200pt]{2.409pt}{0.400pt}}
\put(935.0,233.0){\rule[-0.200pt]{2.409pt}{0.400pt}}
\put(144.0,286.0){\rule[-0.200pt]{4.818pt}{0.400pt}}
\put(126,286){\makebox(0,0)[r]{$10^{-6}$}}
\put(925.0,286.0){\rule[-0.200pt]{4.818pt}{0.400pt}}
\put(144.0,339.0){\rule[-0.200pt]{2.409pt}{0.400pt}}
\put(935.0,339.0){\rule[-0.200pt]{2.409pt}{0.400pt}}
\put(144.0,391.0){\rule[-0.200pt]{2.409pt}{0.400pt}}
\put(935.0,391.0){\rule[-0.200pt]{2.409pt}{0.400pt}}
\put(144.0,444.0){\rule[-0.200pt]{4.818pt}{0.400pt}}
\put(126,444){\makebox(0,0)[r]{$10^{-3}$}}
\put(925.0,444.0){\rule[-0.200pt]{4.818pt}{0.400pt}}
\put(144.0,497.0){\rule[-0.200pt]{2.409pt}{0.400pt}}
\put(935.0,497.0){\rule[-0.200pt]{2.409pt}{0.400pt}}
\put(144.0,550.0){\rule[-0.200pt]{2.409pt}{0.400pt}}
\put(935.0,550.0){\rule[-0.200pt]{2.409pt}{0.400pt}}
\put(144.0,603.0){\rule[-0.200pt]{4.818pt}{0.400pt}}
\put(126,603){\makebox(0,0)[r]{$1$}}
\put(925.0,603.0){\rule[-0.200pt]{4.818pt}{0.400pt}}
\put(152.0,74.0){\rule[-0.200pt]{0.400pt}{2.409pt}}
\put(152.0,593.0){\rule[-0.200pt]{0.400pt}{2.409pt}}
\put(168.0,74.0){\rule[-0.200pt]{0.400pt}{2.409pt}}
\put(168.0,593.0){\rule[-0.200pt]{0.400pt}{2.409pt}}
\put(184.0,74.0){\rule[-0.200pt]{0.400pt}{4.818pt}}
\put(184,37){\makebox(0,0){$-6$}}
\put(184.0,583.0){\rule[-0.200pt]{0.400pt}{4.818pt}}
\put(200.0,74.0){\rule[-0.200pt]{0.400pt}{2.409pt}}
\put(200.0,593.0){\rule[-0.200pt]{0.400pt}{2.409pt}}
\put(216.0,74.0){\rule[-0.200pt]{0.400pt}{2.409pt}}
\put(216.0,593.0){\rule[-0.200pt]{0.400pt}{2.409pt}}
\put(232.0,74.0){\rule[-0.200pt]{0.400pt}{2.409pt}}
\put(232.0,593.0){\rule[-0.200pt]{0.400pt}{2.409pt}}
\put(248.0,74.0){\rule[-0.200pt]{0.400pt}{2.409pt}}
\put(248.0,593.0){\rule[-0.200pt]{0.400pt}{2.409pt}}
\put(264.0,74.0){\rule[-0.200pt]{0.400pt}{4.818pt}}
\put(264,37){\makebox(0,0){$-5$}}
\put(264.0,583.0){\rule[-0.200pt]{0.400pt}{4.818pt}}
\put(280.0,74.0){\rule[-0.200pt]{0.400pt}{2.409pt}}
\put(280.0,593.0){\rule[-0.200pt]{0.400pt}{2.409pt}}
\put(296.0,74.0){\rule[-0.200pt]{0.400pt}{2.409pt}}
\put(296.0,593.0){\rule[-0.200pt]{0.400pt}{2.409pt}}
\put(312.0,74.0){\rule[-0.200pt]{0.400pt}{2.409pt}}
\put(312.0,593.0){\rule[-0.200pt]{0.400pt}{2.409pt}}
\put(328.0,74.0){\rule[-0.200pt]{0.400pt}{2.409pt}}
\put(328.0,593.0){\rule[-0.200pt]{0.400pt}{2.409pt}}
\put(344.0,74.0){\rule[-0.200pt]{0.400pt}{4.818pt}}
\put(344,37){\makebox(0,0){$-4$}}
\put(344.0,583.0){\rule[-0.200pt]{0.400pt}{4.818pt}}
\put(360.0,74.0){\rule[-0.200pt]{0.400pt}{2.409pt}}
\put(360.0,593.0){\rule[-0.200pt]{0.400pt}{2.409pt}}
\put(376.0,74.0){\rule[-0.200pt]{0.400pt}{2.409pt}}
\put(376.0,593.0){\rule[-0.200pt]{0.400pt}{2.409pt}}
\put(392.0,74.0){\rule[-0.200pt]{0.400pt}{2.409pt}}
\put(392.0,593.0){\rule[-0.200pt]{0.400pt}{2.409pt}}
\put(408.0,74.0){\rule[-0.200pt]{0.400pt}{2.409pt}}
\put(408.0,593.0){\rule[-0.200pt]{0.400pt}{2.409pt}}
\put(424.0,74.0){\rule[-0.200pt]{0.400pt}{4.818pt}}
\put(424,37){\makebox(0,0){$-3$}}
\put(424.0,583.0){\rule[-0.200pt]{0.400pt}{4.818pt}}
\put(440.0,74.0){\rule[-0.200pt]{0.400pt}{2.409pt}}
\put(440.0,593.0){\rule[-0.200pt]{0.400pt}{2.409pt}}
\put(456.0,74.0){\rule[-0.200pt]{0.400pt}{2.409pt}}
\put(456.0,593.0){\rule[-0.200pt]{0.400pt}{2.409pt}}
\put(472.0,74.0){\rule[-0.200pt]{0.400pt}{2.409pt}}
\put(472.0,593.0){\rule[-0.200pt]{0.400pt}{2.409pt}}
\put(488.0,74.0){\rule[-0.200pt]{0.400pt}{2.409pt}}
\put(488.0,593.0){\rule[-0.200pt]{0.400pt}{2.409pt}}
\put(504.0,74.0){\rule[-0.200pt]{0.400pt}{4.818pt}}
\put(504,37){\makebox(0,0){$-2$}}
\put(504.0,583.0){\rule[-0.200pt]{0.400pt}{4.818pt}}
\put(520.0,74.0){\rule[-0.200pt]{0.400pt}{2.409pt}}
\put(520.0,593.0){\rule[-0.200pt]{0.400pt}{2.409pt}}
\put(536.0,74.0){\rule[-0.200pt]{0.400pt}{2.409pt}}
\put(536.0,593.0){\rule[-0.200pt]{0.400pt}{2.409pt}}
\put(553.0,74.0){\rule[-0.200pt]{0.400pt}{2.409pt}}
\put(553.0,593.0){\rule[-0.200pt]{0.400pt}{2.409pt}}
\put(569.0,74.0){\rule[-0.200pt]{0.400pt}{2.409pt}}
\put(569.0,593.0){\rule[-0.200pt]{0.400pt}{2.409pt}}
\put(585.0,74.0){\rule[-0.200pt]{0.400pt}{4.818pt}}
\put(585,37){\makebox(0,0){$-1$}}
\put(585.0,583.0){\rule[-0.200pt]{0.400pt}{4.818pt}}
\put(601.0,74.0){\rule[-0.200pt]{0.400pt}{2.409pt}}
\put(601.0,593.0){\rule[-0.200pt]{0.400pt}{2.409pt}}
\put(617.0,74.0){\rule[-0.200pt]{0.400pt}{2.409pt}}
\put(617.0,593.0){\rule[-0.200pt]{0.400pt}{2.409pt}}
\put(633.0,74.0){\rule[-0.200pt]{0.400pt}{2.409pt}}
\put(633.0,593.0){\rule[-0.200pt]{0.400pt}{2.409pt}}
\put(649.0,74.0){\rule[-0.200pt]{0.400pt}{2.409pt}}
\put(649.0,593.0){\rule[-0.200pt]{0.400pt}{2.409pt}}
\put(665.0,74.0){\rule[-0.200pt]{0.400pt}{4.818pt}}
\put(665,37){\makebox(0,0){$0$}}
\put(665.0,583.0){\rule[-0.200pt]{0.400pt}{4.818pt}}
\put(681.0,74.0){\rule[-0.200pt]{0.400pt}{2.409pt}}
\put(681.0,593.0){\rule[-0.200pt]{0.400pt}{2.409pt}}
\put(697.0,74.0){\rule[-0.200pt]{0.400pt}{2.409pt}}
\put(697.0,593.0){\rule[-0.200pt]{0.400pt}{2.409pt}}
\put(713.0,74.0){\rule[-0.200pt]{0.400pt}{2.409pt}}
\put(713.0,593.0){\rule[-0.200pt]{0.400pt}{2.409pt}}
\put(729.0,74.0){\rule[-0.200pt]{0.400pt}{2.409pt}}
\put(729.0,593.0){\rule[-0.200pt]{0.400pt}{2.409pt}}
\put(745.0,74.0){\rule[-0.200pt]{0.400pt}{4.818pt}}
\put(745,37){\makebox(0,0){$1$}}
\put(745.0,583.0){\rule[-0.200pt]{0.400pt}{4.818pt}}
\put(761.0,74.0){\rule[-0.200pt]{0.400pt}{2.409pt}}
\put(761.0,593.0){\rule[-0.200pt]{0.400pt}{2.409pt}}
\put(777.0,74.0){\rule[-0.200pt]{0.400pt}{2.409pt}}
\put(777.0,593.0){\rule[-0.200pt]{0.400pt}{2.409pt}}
\put(793.0,74.0){\rule[-0.200pt]{0.400pt}{2.409pt}}
\put(793.0,593.0){\rule[-0.200pt]{0.400pt}{2.409pt}}
\put(809.0,74.0){\rule[-0.200pt]{0.400pt}{2.409pt}}
\put(809.0,593.0){\rule[-0.200pt]{0.400pt}{2.409pt}}
\put(825.0,74.0){\rule[-0.200pt]{0.400pt}{4.818pt}}
\put(825,37){\makebox(0,0){$2$}}
\put(825.0,583.0){\rule[-0.200pt]{0.400pt}{4.818pt}}
\put(841.0,74.0){\rule[-0.200pt]{0.400pt}{2.409pt}}
\put(841.0,593.0){\rule[-0.200pt]{0.400pt}{2.409pt}}
\put(857.0,74.0){\rule[-0.200pt]{0.400pt}{2.409pt}}
\put(857.0,593.0){\rule[-0.200pt]{0.400pt}{2.409pt}}
\put(873.0,74.0){\rule[-0.200pt]{0.400pt}{2.409pt}}
\put(873.0,593.0){\rule[-0.200pt]{0.400pt}{2.409pt}}
\put(889.0,74.0){\rule[-0.200pt]{0.400pt}{2.409pt}}
\put(889.0,593.0){\rule[-0.200pt]{0.400pt}{2.409pt}}
\put(905.0,74.0){\rule[-0.200pt]{0.400pt}{4.818pt}}
\put(905,37){\makebox(0,0){$3$}}
\put(905.0,583.0){\rule[-0.200pt]{0.400pt}{4.818pt}}
\put(921.0,74.0){\rule[-0.200pt]{0.400pt}{2.409pt}}
\put(921.0,593.0){\rule[-0.200pt]{0.400pt}{2.409pt}}
\put(937.0,74.0){\rule[-0.200pt]{0.400pt}{2.409pt}}
\put(937.0,593.0){\rule[-0.200pt]{0.400pt}{2.409pt}}
\put(144.0,74.0){\rule[-0.200pt]{192.961pt}{0.400pt}}
\put(945.0,74.0){\rule[-0.200pt]{0.400pt}{127.436pt}}
\put(945.0,74.0){\vector(1,0){60}}
\put(985,37){\makebox(0,0){${\mathcal E}$}}
\put(144.0,603.0){\rule[-0.200pt]{192.961pt}{0.400pt}}
\put(144.0,74.0){\rule[-0.200pt]{0.400pt}{127.436pt}}
\put(170,574){\raisebox{-.8pt}{\makebox(0,0){$\Diamond$}}}
\put(280,458){\raisebox{-.8pt}{\makebox(0,0){$\Diamond$}}}
\put(292,521){\raisebox{-.8pt}{\makebox(0,0){$\Diamond$}}}
\put(322,421){\raisebox{-.8pt}{\makebox(0,0){$\Diamond$}}}
\put(353,465){\raisebox{-.8pt}{\makebox(0,0){$\Diamond$}}}
\put(364,370){\raisebox{-.8pt}{\makebox(0,0){$\Diamond$}}}
\put(377,456){\raisebox{-.8pt}{\makebox(0,0){$\Diamond$}}}
\put(392,390){\raisebox{-.8pt}{\makebox(0,0){$\Diamond$}}}
\put(400,402){\raisebox{-.8pt}{\makebox(0,0){$\Diamond$}}}
\put(404,438){\raisebox{-.8pt}{\makebox(0,0){$\Diamond$}}}
\put(425,392){\raisebox{-.8pt}{\makebox(0,0){$\Diamond$}}}
\put(895,98){\raisebox{-.8pt}{\makebox(0,0){$\Diamond$}}}
\put(438,363){\raisebox{-.8pt}{\makebox(0,0){$\Diamond$}}}
\put(885,149){\raisebox{-.8pt}{\makebox(0,0){$\Diamond$}}}
\put(446,372){\raisebox{-.8pt}{\makebox(0,0){$\Diamond$}}}
\put(455,391){\raisebox{-.8pt}{\makebox(0,0){$\Diamond$}}}
\put(459,381){\raisebox{-.8pt}{\makebox(0,0){$\Diamond$}}}
\put(466,423){\raisebox{-.8pt}{\makebox(0,0){$\Diamond$}}}
\put(473,362){\raisebox{-.8pt}{\makebox(0,0){$\Diamond$}}}
\put(480,298){\raisebox{-.8pt}{\makebox(0,0){$\Diamond$}}}
\put(486,367){\raisebox{-.8pt}{\makebox(0,0){$\Diamond$}}}
\put(842,154){\raisebox{-.8pt}{\makebox(0,0){$\Diamond$}}}
\put(492,366){\raisebox{-.8pt}{\makebox(0,0){$\Diamond$}}}
\put(835,149){\raisebox{-.8pt}{\makebox(0,0){$\Diamond$}}}
\put(500,326){\raisebox{-.8pt}{\makebox(0,0){$\Diamond$}}}
\put(511,341){\raisebox{-.8pt}{\makebox(0,0){$\Diamond$}}}
\put(512,336){\raisebox{-.8pt}{\makebox(0,0){$\Diamond$}}}
\put(809,172){\raisebox{-.8pt}{\makebox(0,0){$\Diamond$}}}
\put(523,361){\raisebox{-.8pt}{\makebox(0,0){$\Diamond$}}}
\put(801,198){\raisebox{-.8pt}{\makebox(0,0){$\Diamond$}}}
\put(531,280){\raisebox{-.8pt}{\makebox(0,0){$\Diamond$}}}
\put(539,292){\raisebox{-.8pt}{\makebox(0,0){$\Diamond$}}}
\put(543,352){\raisebox{-.8pt}{\makebox(0,0){$\Diamond$}}}
\put(552,355){\raisebox{-.8pt}{\makebox(0,0){$\Diamond$}}}
\put(556,360){\raisebox{-.8pt}{\makebox(0,0){$\Diamond$}}}
\put(771,92){\raisebox{-.8pt}{\makebox(0,0){$\Diamond$}}}
\put(564,280){\raisebox{-.8pt}{\makebox(0,0){$\Diamond$}}}
\put(573,300){\raisebox{-.8pt}{\makebox(0,0){$\Diamond$}}}
\put(754,192){\raisebox{-.8pt}{\makebox(0,0){$\Diamond$}}}
\put(579,302){\raisebox{-.8pt}{\makebox(0,0){$\Diamond$}}}
\put(746,199){\raisebox{-.8pt}{\makebox(0,0){$\Diamond$}}}
\put(584,343){\raisebox{-.8pt}{\makebox(0,0){$\Diamond$}}}
\put(739,163){\raisebox{-.8pt}{\makebox(0,0){$\Diamond$}}}
\put(592,291){\raisebox{-.8pt}{\makebox(0,0){$\Diamond$}}}
\put(734,264){\raisebox{-.8pt}{\makebox(0,0){$\Diamond$}}}
\put(600,255){\raisebox{-.8pt}{\makebox(0,0){$\Diamond$}}}
\put(604,198){\raisebox{-.8pt}{\makebox(0,0){$\Diamond$}}}
\put(722,186){\raisebox{-.8pt}{\makebox(0,0){$\Diamond$}}}
\put(612,309){\raisebox{-.8pt}{\makebox(0,0){$\Diamond$}}}
\put(617,288){\raisebox{-.8pt}{\makebox(0,0){$\Diamond$}}}
\put(707,240){\raisebox{-.8pt}{\makebox(0,0){$\Diamond$}}}
\put(629,268){\raisebox{-.8pt}{\makebox(0,0){$\Diamond$}}}
\put(691,228){\raisebox{-.8pt}{\makebox(0,0){$\Diamond$}}}
\put(687,297){\raisebox{-.8pt}{\makebox(0,0){$\Diamond$}}}
\put(647,287){\raisebox{-.8pt}{\makebox(0,0){$\Diamond$}}}
\put(681,268){\raisebox{-.8pt}{\makebox(0,0){$\Diamond$}}}
\put(654,272){\raisebox{-.8pt}{\makebox(0,0){$\Diamond$}}}
\put(655,263){\raisebox{-.8pt}{\makebox(0,0){$\Diamond$}}}
\put(673,218){\raisebox{-.8pt}{\makebox(0,0){$\Diamond$}}}
\put(664,270){\raisebox{-.8pt}{\makebox(0,0){$\Diamond$}}}
\put(144.00,563.00){\usebox{\plotpoint}}
\put(161.60,552.00){\usebox{\plotpoint}}
\put(179.62,541.74){\usebox{\plotpoint}}
\put(197.49,531.20){\usebox{\plotpoint}}
\put(215.09,520.20){\usebox{\plotpoint}}
\put(233.10,509.94){\usebox{\plotpoint}}
\put(250.70,498.94){\usebox{\plotpoint}}
\put(268.30,487.93){\usebox{\plotpoint}}
\put(286.40,477.80){\usebox{\plotpoint}}
\put(304.19,467.13){\usebox{\plotpoint}}
\put(321.79,456.13){\usebox{\plotpoint}}
\put(339.81,445.87){\usebox{\plotpoint}}
\put(357.41,434.87){\usebox{\plotpoint}}
\put(375.28,424.33){\usebox{\plotpoint}}
\put(393.20,413.90){\usebox{\plotpoint}}
\put(410.89,403.07){\usebox{\plotpoint}}
\put(428.49,392.07){\usebox{\plotpoint}}
\put(446.26,381.37){\usebox{\plotpoint}}
\put(464.38,371.26){\usebox{\plotpoint}}
\put(481.98,360.26){\usebox{\plotpoint}}
\put(500.00,350.00){\usebox{\plotpoint}}
\put(517.60,339.00){\usebox{\plotpoint}}
\put(535.20,328.00){\usebox{\plotpoint}}
\put(553.29,317.85){\usebox{\plotpoint}}
\put(571.08,307.20){\usebox{\plotpoint}}
\put(588.68,296.20){\usebox{\plotpoint}}
\put(606.36,285.32){\usebox{\plotpoint}}
\put(624.30,274.94){\usebox{\plotpoint}}
\put(642.17,264.39){\usebox{\plotpoint}}
\put(660.09,253.96){\usebox{\plotpoint}}
\put(677.79,243.13){\usebox{\plotpoint}}
\put(695.39,232.13){\usebox{\plotpoint}}
\put(713.15,221.42){\usebox{\plotpoint}}
\put(731.27,211.33){\usebox{\plotpoint}}
\put(748.87,200.33){\usebox{\plotpoint}}
\put(766.48,189.33){\usebox{\plotpoint}}
\put(784.49,179.07){\usebox{\plotpoint}}
\put(802.09,168.07){\usebox{\plotpoint}}
\put(820.18,157.91){\usebox{\plotpoint}}
\put(837.98,147.26){\usebox{\plotpoint}}
\put(855.58,136.26){\usebox{\plotpoint}}
\put(873.24,125.38){\usebox{\plotpoint}}
\put(891.19,115.00){\usebox{\plotpoint}}
\put(909.06,104.46){\usebox{\plotpoint}}
\put(926.67,93.46){\usebox{\plotpoint}}
\put(944.68,83.20){\usebox{\plotpoint}}
\put(945.00,83.00){\usebox{\plotpoint}}
\end{picture}
\caption{The N\'eel probability $p_{\tN}$ versus energy $\mathcal E$ for states
with $S=4$, $\Gamma=B_1$.\label{n04b1}}
\end{figure}

\begin{figure}
\setlength{\unitlength}{0.240900pt}
\ifx\plotpoint\undefined\newsavebox{\plotpoint}\fi
\sbox{\plotpoint}{\rule[-0.200pt]{0.400pt}{0.400pt}}%
\begin{picture}(1005,603)(0,0)
\font\gnuplot=cmr10 at 9pt
\gnuplot
\sbox{\plotpoint}{\rule[-0.200pt]{0.400pt}{0.400pt}}%
\put(127.0,66.0){\rule[-0.200pt]{2.409pt}{0.400pt}}
\put(127.0,83.0){\rule[-0.200pt]{2.409pt}{0.400pt}}
\put(127.0,100.0){\rule[-0.200pt]{2.409pt}{0.400pt}}
\put(109,117){\makebox(0,0)[r]{$-6$}}
\put(127.0,117.0){\rule[-0.200pt]{4.818pt}{0.400pt}}
\put(127.0,134.0){\rule[-0.200pt]{2.409pt}{0.400pt}}
\put(127.0,151.0){\rule[-0.200pt]{2.409pt}{0.400pt}}
\put(127.0,168.0){\rule[-0.200pt]{2.409pt}{0.400pt}}
\put(127.0,185.0){\rule[-0.200pt]{2.409pt}{0.400pt}}
\put(109,202){\makebox(0,0)[r]{$-5$}}
\put(127.0,202.0){\rule[-0.200pt]{4.818pt}{0.400pt}}
\put(127.0,219.0){\rule[-0.200pt]{2.409pt}{0.400pt}}
\put(127.0,236.0){\rule[-0.200pt]{2.409pt}{0.400pt}}
\put(127.0,253.0){\rule[-0.200pt]{2.409pt}{0.400pt}}
\put(127.0,270.0){\rule[-0.200pt]{2.409pt}{0.400pt}}
\put(109,287){\makebox(0,0)[r]{$-4$}}
\put(127.0,287.0){\rule[-0.200pt]{4.818pt}{0.400pt}}
\put(127.0,304.0){\rule[-0.200pt]{2.409pt}{0.400pt}}
\put(127.0,321.0){\rule[-0.200pt]{2.409pt}{0.400pt}}
\put(127.0,339.0){\rule[-0.200pt]{2.409pt}{0.400pt}}
\put(127.0,356.0){\rule[-0.200pt]{2.409pt}{0.400pt}}
\put(109,373){\makebox(0,0)[r]{$-3$}}
\put(127.0,373.0){\rule[-0.200pt]{4.818pt}{0.400pt}}
\put(127.0,390.0){\rule[-0.200pt]{2.409pt}{0.400pt}}
\put(127.0,407.0){\rule[-0.200pt]{2.409pt}{0.400pt}}
\put(127.0,424.0){\rule[-0.200pt]{2.409pt}{0.400pt}}
\put(127.0,441.0){\rule[-0.200pt]{2.409pt}{0.400pt}}
\put(109,458){\makebox(0,0)[r]{$-2$}}
\put(127.0,458.0){\rule[-0.200pt]{4.818pt}{0.400pt}}
\put(127.0,475.0){\rule[-0.200pt]{2.409pt}{0.400pt}}
\put(127.0,492.0){\rule[-0.200pt]{2.409pt}{0.400pt}}
\put(127.0,509.0){\rule[-0.200pt]{2.409pt}{0.400pt}}
\put(127.0,526.0){\rule[-0.200pt]{2.409pt}{0.400pt}}
\put(109,543){\makebox(0,0)[r]{$-1$}}
\put(127.0,543.0){\rule[-0.200pt]{4.818pt}{0.400pt}}
\put(127.0,560.0){\rule[-0.200pt]{2.409pt}{0.400pt}}
\put(127.0,577.0){\rule[-0.200pt]{2.409pt}{0.400pt}}
\put(153.0,49.0){\rule[-0.200pt]{0.400pt}{4.818pt}}
\put(153,13){\makebox(0,0){$0$}}
\put(153.0,566.0){\rule[-0.200pt]{0.400pt}{4.818pt}}
\put(205.0,49.0){\rule[-0.200pt]{0.400pt}{2.409pt}}
\put(205.0,576.0){\rule[-0.200pt]{0.400pt}{2.409pt}}
\put(258.0,49.0){\rule[-0.200pt]{0.400pt}{4.818pt}}
\put(258,13){\makebox(0,0){$2$}}
\put(258.0,566.0){\rule[-0.200pt]{0.400pt}{4.818pt}}
\put(310.0,49.0){\rule[-0.200pt]{0.400pt}{2.409pt}}
\put(310.0,576.0){\rule[-0.200pt]{0.400pt}{2.409pt}}
\put(362.0,49.0){\rule[-0.200pt]{0.400pt}{4.818pt}}
\put(362,13){\makebox(0,0){$4$}}
\put(362.0,566.0){\rule[-0.200pt]{0.400pt}{4.818pt}}
\put(415.0,49.0){\rule[-0.200pt]{0.400pt}{2.409pt}}
\put(415.0,576.0){\rule[-0.200pt]{0.400pt}{2.409pt}}
\put(467.0,49.0){\rule[-0.200pt]{0.400pt}{4.818pt}}
\put(467,13){\makebox(0,0){$6$}}
\put(467.0,566.0){\rule[-0.200pt]{0.400pt}{4.818pt}}
\put(519.0,49.0){\rule[-0.200pt]{0.400pt}{2.409pt}}
\put(519.0,576.0){\rule[-0.200pt]{0.400pt}{2.409pt}}
\put(571.0,49.0){\rule[-0.200pt]{0.400pt}{4.818pt}}
\put(571,13){\makebox(0,0){$8$}}
\put(571.0,566.0){\rule[-0.200pt]{0.400pt}{4.818pt}}
\put(624.0,49.0){\rule[-0.200pt]{0.400pt}{2.409pt}}
\put(624.0,576.0){\rule[-0.200pt]{0.400pt}{2.409pt}}
\put(676.0,49.0){\rule[-0.200pt]{0.400pt}{4.818pt}}
\put(676,13){\makebox(0,0){$10$}}
\put(676.0,566.0){\rule[-0.200pt]{0.400pt}{4.818pt}}
\put(728.0,49.0){\rule[-0.200pt]{0.400pt}{2.409pt}}
\put(728.0,576.0){\rule[-0.200pt]{0.400pt}{2.409pt}}
\put(781.0,49.0){\rule[-0.200pt]{0.400pt}{4.818pt}}
\put(781,13){\makebox(0,0){$12$}}
\put(781.0,566.0){\rule[-0.200pt]{0.400pt}{4.818pt}}
\put(833.0,49.0){\rule[-0.200pt]{0.400pt}{2.409pt}}
\put(833.0,576.0){\rule[-0.200pt]{0.400pt}{2.409pt}}
\put(877,86){\makebox(0,0)[l]{$-15$}}
\put(849.0,62.0){\rule[-0.200pt]{2.409pt}{0.400pt}}
\put(839.0,86.0){\rule[-0.200pt]{4.818pt}{0.400pt}}
\put(849.0,111.0){\rule[-0.200pt]{2.409pt}{0.400pt}}
\put(849.0,135.0){\rule[-0.200pt]{2.409pt}{0.400pt}}
\put(849.0,159.0){\rule[-0.200pt]{2.409pt}{0.400pt}}
\put(849.0,184.0){\rule[-0.200pt]{2.409pt}{0.400pt}}
\put(877,208){\makebox(0,0)[l]{$-10$}}
\put(839.0,208.0){\rule[-0.200pt]{4.818pt}{0.400pt}}
\put(849.0,232.0){\rule[-0.200pt]{2.409pt}{0.400pt}}
\put(849.0,257.0){\rule[-0.200pt]{2.409pt}{0.400pt}}
\put(849.0,281.0){\rule[-0.200pt]{2.409pt}{0.400pt}}
\put(849.0,306.0){\rule[-0.200pt]{2.409pt}{0.400pt}}
\put(877,330){\makebox(0,0)[l]{$-5$}}
\put(839.0,330.0){\rule[-0.200pt]{4.818pt}{0.400pt}}
\put(849.0,354.0){\rule[-0.200pt]{2.409pt}{0.400pt}}
\put(849.0,379.0){\rule[-0.200pt]{2.409pt}{0.400pt}}
\put(849.0,403.0){\rule[-0.200pt]{2.409pt}{0.400pt}}
\put(849.0,428.0){\rule[-0.200pt]{2.409pt}{0.400pt}}
\put(877,452){\makebox(0,0)[l]{0}}
\put(839.0,452.0){\rule[-0.200pt]{4.818pt}{0.400pt}}
\put(849.0,476.0){\rule[-0.200pt]{2.409pt}{0.400pt}}
\put(849.0,501.0){\rule[-0.200pt]{2.409pt}{0.400pt}}
\put(849.0,525.0){\rule[-0.200pt]{2.409pt}{0.400pt}}
\put(849.0,549.0){\rule[-0.200pt]{2.409pt}{0.400pt}}
\put(877,574){\makebox(0,0)[l]{5}}
\put(839.0,574.0){\rule[-0.200pt]{4.818pt}{0.400pt}}
\put(127.0,49.0){\rule[-0.200pt]{176.339pt}{0.400pt}}
\put(859,49){\vector(1,0){80}}
\put(879,13){\makebox(0,0)[l]{$S$}}
\put(859.0,49.0){\rule[-0.200pt]{0.400pt}{129.364pt}}
\put(127.0,586.0){\rule[-0.200pt]{176.339pt}{0.400pt}}
\put(127.0,49.0){\rule[-0.200pt]{0.400pt}{129.364pt}}
\put(30,506){\makebox(0,0){$\alpha(S)$}}
\put(940,506){\makebox(0,0){$\beta(S)$}}
\put(153,514){\raisebox{-.8pt}{\makebox(0,0){$\Box$}}}
\put(205,504){\raisebox{-.8pt}{\makebox(0,0){$\Box$}}}
\put(258,460){\raisebox{-.8pt}{\makebox(0,0){$\Box$}}}
\put(310,449){\raisebox{-.8pt}{\makebox(0,0){$\Box$}}}
\put(362,443){\raisebox{-.8pt}{\makebox(0,0){$\Box$}}}
\put(415,437){\raisebox{-.8pt}{\makebox(0,0){$\Box$}}}
\put(467,432){\raisebox{-.8pt}{\makebox(0,0){$\Box$}}}
\put(519,424){\raisebox{-.8pt}{\makebox(0,0){$\Box$}}}
\put(571,413){\raisebox{-.8pt}{\makebox(0,0){$\Box$}}}
\put(624,395){\raisebox{-.8pt}{\makebox(0,0){$\Box$}}}
\put(676,372){\raisebox{-.8pt}{\makebox(0,0){$\Box$}}}
\put(728,332){\raisebox{-.8pt}{\makebox(0,0){$\Box$}}}
\put(781,278){\raisebox{-.8pt}{\makebox(0,0){$\Box$}}}
\put(833,145){\raisebox{-.8pt}{\makebox(0,0){$\Box$}}}
\sbox{\plotpoint}{\rule[-0.500pt]{1.000pt}{1.000pt}}%
\put(147.76,447.00){\usebox{\plotpoint}}
\put(168.51,447.00){\usebox{\plotpoint}}
\put(189.27,447.00){\usebox{\plotpoint}}
\put(209.95,446.00){\usebox{\plotpoint}}
\put(230.71,446.00){\usebox{\plotpoint}}
\put(251.46,446.00){\usebox{\plotpoint}}
\put(272.22,446.00){\usebox{\plotpoint}}
\put(292.90,445.00){\usebox{\plotpoint}}
\put(313.66,445.00){\usebox{\plotpoint}}
\put(334.35,444.00){\usebox{\plotpoint}}
\put(355.04,443.14){\usebox{\plotpoint}}
\put(375.74,442.32){\usebox{\plotpoint}}
\put(396.45,441.57){\usebox{\plotpoint}}
\put(417.10,440.00){\usebox{\plotpoint}}
\put(437.78,439.00){\usebox{\plotpoint}}
\put(458.39,437.00){\usebox{\plotpoint}}
\put(478.98,434.43){\usebox{\plotpoint}}
\put(499.61,432.63){\usebox{\plotpoint}}
\put(520.00,428.86){\usebox{\plotpoint}}
\put(540.37,425.18){\usebox{\plotpoint}}
\put(560.74,421.32){\usebox{\plotpoint}}
\put(580.83,416.19){\usebox{\plotpoint}}
\put(600.58,409.86){\usebox{\plotpoint}}
\put(620.05,402.84){\usebox{\plotpoint}}
\put(639.21,394.90){\usebox{\plotpoint}}
\put(657.40,384.91){\usebox{\plotpoint}}
\put(675.16,374.28){\usebox{\plotpoint}}
\put(692.32,362.63){\usebox{\plotpoint}}
\put(708.41,349.59){\usebox{\plotpoint}}
\put(723.67,335.66){\usebox{\plotpoint}}
\put(737.86,320.53){\usebox{\plotpoint}}
\put(751.10,304.57){\usebox{\plotpoint}}
\put(763.62,288.03){\usebox{\plotpoint}}
\put(775.21,270.83){\usebox{\plotpoint}}
\put(786.01,253.13){\usebox{\plotpoint}}
\put(796.03,234.95){\usebox{\plotpoint}}
\put(805.39,216.45){\usebox{\plotpoint}}
\put(814.58,197.84){\usebox{\plotpoint}}
\put(822.52,178.67){\usebox{\plotpoint}}
\put(830.07,159.33){\usebox{\plotpoint}}
\put(837.69,140.03){\usebox{\plotpoint}}
\multiput(844,122)(7.093,-19.506){2}{\usebox{\plotpoint}}
\put(147.76,96.00){\usebox{\plotpoint}}
\put(168.51,96.00){\usebox{\plotpoint}}
\put(189.24,96.41){\usebox{\plotpoint}}
\put(209.96,97.00){\usebox{\plotpoint}}
\put(230.72,97.00){\usebox{\plotpoint}}
\put(251.42,97.80){\usebox{\plotpoint}}
\put(272.16,98.00){\usebox{\plotpoint}}
\put(292.86,99.00){\usebox{\plotpoint}}
\put(313.55,100.00){\usebox{\plotpoint}}
\put(334.23,101.00){\usebox{\plotpoint}}
\put(354.93,101.85){\usebox{\plotpoint}}
\put(375.60,103.00){\usebox{\plotpoint}}
\put(396.27,104.41){\usebox{\plotpoint}}
\put(416.90,106.24){\usebox{\plotpoint}}
\put(437.48,108.94){\usebox{\plotpoint}}
\put(458.05,111.76){\usebox{\plotpoint}}
\put(478.51,115.00){\usebox{\plotpoint}}
\put(498.79,119.26){\usebox{\plotpoint}}
\put(518.97,123.99){\usebox{\plotpoint}}
\put(539.00,129.43){\usebox{\plotpoint}}
\put(558.62,136.12){\usebox{\plotpoint}}
\put(577.84,143.93){\usebox{\plotpoint}}
\put(596.70,152.59){\usebox{\plotpoint}}
\put(614.69,162.78){\usebox{\plotpoint}}
\put(632.34,173.67){\usebox{\plotpoint}}
\put(648.82,186.27){\usebox{\plotpoint}}
\put(664.53,199.84){\usebox{\plotpoint}}
\put(679.35,214.35){\usebox{\plotpoint}}
\put(692.92,230.04){\usebox{\plotpoint}}
\put(706.12,246.03){\usebox{\plotpoint}}
\put(718.20,262.90){\usebox{\plotpoint}}
\put(729.02,280.60){\usebox{\plotpoint}}
\put(739.47,298.52){\usebox{\plotpoint}}
\put(749.30,316.79){\usebox{\plotpoint}}
\put(758.26,335.51){\usebox{\plotpoint}}
\put(766.86,354.38){\usebox{\plotpoint}}
\put(775.08,373.44){\usebox{\plotpoint}}
\put(782.69,392.74){\usebox{\plotpoint}}
\put(789.45,412.36){\usebox{\plotpoint}}
\put(796.34,431.94){\usebox{\plotpoint}}
\put(802.93,451.61){\usebox{\plotpoint}}
\put(809.06,471.44){\usebox{\plotpoint}}
\multiput(815,490)(5.209,20.091){2}{\usebox{\plotpoint}}
\put(825.61,531.44){\usebox{\plotpoint}}
\multiput(829,545)(5.186,20.097){2}{\usebox{\plotpoint}}
\put(153,158){\raisebox{-.8pt}{\makebox(0,0){$\Diamond$}}}
\put(205,164){\raisebox{-.8pt}{\makebox(0,0){$\Diamond$}}}
\put(258,91){\raisebox{-.8pt}{\makebox(0,0){$\Diamond$}}}
\put(310,86){\raisebox{-.8pt}{\makebox(0,0){$\Diamond$}}}
\put(362,94){\raisebox{-.8pt}{\makebox(0,0){$\Diamond$}}}
\put(415,105){\raisebox{-.8pt}{\makebox(0,0){$\Diamond$}}}
\put(467,119){\raisebox{-.8pt}{\makebox(0,0){$\Diamond$}}}
\put(519,136){\raisebox{-.8pt}{\makebox(0,0){$\Diamond$}}}
\put(571,156){\raisebox{-.8pt}{\makebox(0,0){$\Diamond$}}}
\put(624,180){\raisebox{-.8pt}{\makebox(0,0){$\Diamond$}}}
\put(676,216){\raisebox{-.8pt}{\makebox(0,0){$\Diamond$}}}
\put(728,271){\raisebox{-.8pt}{\makebox(0,0){$\Diamond$}}}
\put(781,364){\raisebox{-.8pt}{\makebox(0,0){$\Diamond$}}}
\put(833,572){\raisebox{-.8pt}{\makebox(0,0){$\Diamond$}}}
\end{picture}
\caption{Coefficients $\alpha(S)$ (squares) and $\beta(S)$ (diamonds) of estimations
$p_{\tN}=\exp(\alpha(S){\mathcal E}+\beta(S))$ for $0\le S\le13$ and the best-fit
exponential lines (dotted).\label{npabx}}
\end{figure}
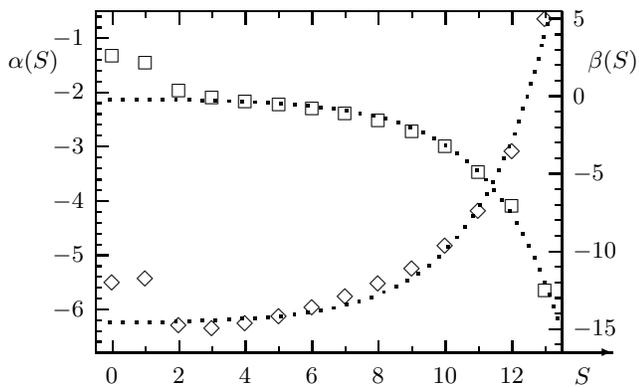

To have a bit deeper insight we have analyzed the N\'eel probability $p_{\tN}$
for other $S$-multiplets. As it has been mentioned earlier only states with 
$S$ even, $\Gamma=B_1$ and $\Gamma=A_1$ for $S$ odd are taken into account.
The lowest lying state (with energy ${\mathcal E}_0(S)$) has always
the maximum $p_{\tN}$, much higher than in the other states. A rough estimate
shows exponential decreasing. For example, in the case $S=4$, $\Gamma=B_1$ the best-fit line 
is given as $p_{\tN}({\mathcal E})=\exp({-2.176}\,{\mathcal E}-14.700)$ 
(see also Fig.~\ref{n04b1}).  Fitting a function
$\exp(\alpha(S){\mathcal E}+\beta(S))$ to $p_{\tN}$ in each $S$-multiplet we have
obtained series $\alpha(S)$, $\beta(S)$ which can be approximated by exponential
functions $\alpha(S)={-\exp}(0.463S-4.781)-2.122$ and $\beta(S)=\exp(0.463S-3.061)-14.643$, 
respectively (see Fig.~\ref{npabx}). Points for $S=0,1$ have not been taking into account
due to their irregular behavior.

All above considerations concern the N\'eel probability $p_{\tN}$. However, as it has
been mentioned discussing the ground state, the N\'eel states $\ket{\tN1}$ and $\ket{\tN2}$
are only two of $ns+1$ possible states $\ket{S_A,m;S_B,{-m}}$ with $S_A=S_B={1\over2}ns$ and 
$|m|\le {1\over2}ns$, where $S_A$ ($S_B$) denotes the total spin in one sublattice of a bipartite
antiferromagnet. Coupling of sublattice spins leads to $(ns+1)^2$ states with the total
spin $0\le S\le ns$. In each $S$-multiplet there is one state $\ket{S,0}$ with the
magnetization equal to zero. This state is a linear combination of $ns+1$ states
$\ket{S_A,m;S_B,{-m}}$, namely
 \[
  \ket{S,0}=\sum_{m=-ns/2}^{ns/2} \left[\begin{array}{ccc}
  {\textstyle{1\over2}}ns&{\textstyle{1\over2}}ns&S\\m&-m&0\end{array}\right]
    \ket{{\textstyle{1\over2}}ns,m;{\textstyle{1\over2}}ns,{-m}}\,,
 \] 
 where the expression in brackets is the Clebsch--Gordan coefficient (CGC) for coupling
spin representations $D^s$.\cite{edm} In the case $S=0$ the appropriate CGC's
are equal to 
 \[(-1)^{{1\over2}ns-m}(ns+1)^{1\over2}\,,\]
 so all states 
$\ket{{1\over2}ns,m;{1\over2}ns,{-m}}$ have the same 
probability and, therefore, the global N\'eel probability $p^*_{\tN}=(ns+1)p_{\tN}$. Note
the N\'eel states $\ket{\tN1}$ and $\ket{\tN2}$ correspond to $m=\pm {1\over2}ns$, respectively.
Hence, these states enter $\ket{0,0}$ with opposite signs for odd $ns$, according with
the Marshall rule.\cite{marsh,casp} 

In this paper we consider states
with total magnetization equal to zero and the N\'eel states correspond to sublattice states
with $S_A=S_B=|m|=ns/2$. So, we are interested in the CGC's for these values; they
can be derived from a general formula\cite{edm} and one obtains 
\begin{eqnarray}\nonumber
  \alpha_{\tN1(2)}&=&\left[\begin{array}{ccc}
  {1\over2}ns&{1\over2}ns&S\\\pm{1\over2} ns&\mp{1\over2} ns&0\end{array}\right] \\
 &=& (\pm1)^{ns+S}  
\frac{(ns)!\sqrt{2S+1}}{[(ns-S)!(ns+S+1)!]^{1\over2}}\,.
\label{cgc}
\end{eqnarray}
  For example, the maximum total spin $S=ns$ yields equal coefficients for the both N\'eel
states 
  \[ 
     \alpha_{\tN1}=\alpha_{\tN2}=\frac{(ns)!}{\sqrt{(2ns)!}}\,;
  \]
 this state always has the symmetry $A_1$ ($\Gamma_0$ in a general case).
 To obtain the global N\'eel probability we have to divide $p_{\tN1(2)}$
 by $\alpha_{\tN1}^2$. The results for $S=14$ 
and $S=15$ 
can be easily predicted. In both cases there is the unique state with these labels and $M=0$,
so we have to obtain $|a_{\tN1}|=\sqrt{2}|\alpha_{\tN1}|$. In fact, our calculations 
confirm this relation, so in these cases $p^*_{\tN}=1$. The global N\'eel probabilities
for the states with the lowest energy ${\mathcal E}_0(S)$ in each $S$-multiplet are 
collected in Table~\ref{npt}.

\begin{table}
\caption{The N\'eel probability $p_{\tN1}$, the CGC for the N\'eel state in $\ket{S,0}$,
and the global N\'eel probability $p^*_{\tN}=\alpha_{\tN1}^{-2}p_{\tN1}$ for the lowest
lying states in each $S$-multiplet.\label{npt}}

\begin{ruledtabular} \begin{tabular}{rdrd}
  $S$ & \multicolumn{1}{c}{$10^3\cdot p_{\tN1} [\%]$}  & $\alpha_{\tN1}^{-2}$   &  
\multicolumn{1}{c}{$p^*_{\tN} [\%]$}  \\ \hline
  0 &  5391.126       & 16          &  86.258 \\
  1 & 14305.617       & 272/45      &  86.470 \\
  2 & 18634.880       & 816/175     &  86.892 \\
  3 & 17979.305       & 15504/3185  &  87.520 \\
  4 & 13999.674       & 5168/819    &  88.340 \\
  5 &  9062.557       & 15504/1573  &  89.324 \\
  6 &  4928.709       & 15504/845   &  90.432 \\
  7 &  2254.612       & 118864/2925 &  91.621 \\
  8 &   863.2259      & 6992/65     &  92.857 \\
  9 &   273.7878      & 31280/91    &  94.111 \\
 10 &    70.76512     & 594320/441  &  95.368 \\
 11 &    14.54008     & 46512/7     &  96.613 \\
 12 &     2.286329    & 1069776/25  &  97.834 \\
 13 &     0.2584271   & 3447056/9   &  98.979 \\
 14 &     0.0186955   & 5348880     & 100.000 \\
 15 &     0.00064467  & 155117520   & 100.000 \\
\end{tabular} \end{ruledtabular} \end{table}

Assuming (as Iske and Caspers in Ref.~\onlinecite{casp}) 
that the global N\'eel probability is related to the long-range ordering
we come to a conclusion that it is present in the states $\ket{ns,0}$
and $\ket{ns-1,0}$. In the other states $\ket{S,0}$ this ordering is slightly `broken' 
or, in the other words, the ground state of quantum antiferromagnetic rings are 
ordered to some extent only. In our case it is reflected by the 
value $p^*_{\tN}=0.86258$. However, this number has to be compared 
with $p^*_{\tN,\text{dis}}$ in a completely disordered state.\cite{casp} One of possible
definitions of such a state, appropriate in our considerations, is to combine   
all vectors with $M=0$ and $\Gamma=B_1$. There are 385 such vectors, 
so $a_{\tN}=385^{-{1\over2}}\approx0.051$; 
then $p_{\tN,\text{dis}}^*\approx0.021$. Much larger global N\'eel probability 
we obtain taking into account only states with $S=0$. There are twenty such states
and we take a sum of N\'eel orbit coefficients divided by $\sqrt{20}$ as $a_{\tN}$ is 
this state. In this approach $p_{\tN,\text{dis}}^*\approx0.235$. The value obtained 
is almost four times smaller than that in the ground state. Of course, this ratio should
decrease for larger rings, since---as it is well known---no long-order exists in
linear antiferromagnets. On the other hand, however, small antiferromagnetic nanoclusters 
should exhibit a well-ordered ground state.

\section{Final remarks}\label{diss}

\begin{figure}
\setlength{\unitlength}{0.240900pt}
\ifx\plotpoint\undefined\newsavebox{\plotpoint}\fi
\begin{picture}(1020,603)(10,0)
\font\gnuplot=cmr10 at 9pt
\gnuplot
\sbox{\plotpoint}{\rule[-0.200pt]{0.600pt}{0.600pt}}%
\put(126.0, 74.0){\rule[-0.200pt]{2.409pt}{0.400pt}}
\put(935.0, 74.0){\rule[-0.200pt]{2.409pt}{0.400pt}}
\put(126.0, 74.0){\rule[-0.200pt]{4.818pt}{0.400pt}}
\put(925.0, 74.0){\rule[-0.200pt]{4.818pt}{0.400pt}}
\put(126.0,105.0){\rule[-0.200pt]{2.409pt}{0.400pt}}
\put(935.0,105.0){\rule[-0.200pt]{2.409pt}{0.400pt}}
\put(126.0,136.0){\rule[-0.200pt]{2.409pt}{0.400pt}}
\put(935.0,136.0){\rule[-0.200pt]{2.409pt}{0.400pt}}
\put(126.0,166.0){\rule[-0.200pt]{2.409pt}{0.400pt}}
\put(935.0,166.0){\rule[-0.200pt]{2.409pt}{0.400pt}}
\put(126.0,197.0){\rule[-0.200pt]{2.409pt}{0.400pt}}
\put(935.0,197.0){\rule[-0.200pt]{2.409pt}{0.400pt}}
\put(126.0,228.0){\rule[-0.200pt]{2.409pt}{0.400pt}}
\put(935.0,228.0){\rule[-0.200pt]{2.409pt}{0.400pt}}
\put(126.0,228.0){\rule[-0.200pt]{4.818pt}{0.400pt}}
\put(925.0,228.0){\rule[-0.200pt]{4.818pt}{0.400pt}}
\put(126.0,259.0){\rule[-0.200pt]{2.409pt}{0.400pt}}
\put(935.0,259.0){\rule[-0.200pt]{2.409pt}{0.400pt}}
\put(126.0,289.0){\rule[-0.200pt]{2.409pt}{0.400pt}}
\put(935.0,289.0){\rule[-0.200pt]{2.409pt}{0.400pt}}
\put(126.0,320.0){\rule[-0.200pt]{2.409pt}{0.400pt}}
\put(935.0,320.0){\rule[-0.200pt]{2.409pt}{0.400pt}}
\put(126.0,351.0){\rule[-0.200pt]{2.409pt}{0.400pt}}
\put(935.0,351.0){\rule[-0.200pt]{2.409pt}{0.400pt}}
\put(126.0,382.0){\rule[-0.200pt]{2.409pt}{0.400pt}}
\put(935.0,382.0){\rule[-0.200pt]{2.409pt}{0.400pt}}
\put(126.0,382.0){\rule[-0.200pt]{4.818pt}{0.400pt}}
\put(925.0,382.0){\rule[-0.200pt]{4.818pt}{0.400pt}}
\put(126.0,412.0){\rule[-0.200pt]{2.409pt}{0.400pt}}
\put(935.0,412.0){\rule[-0.200pt]{2.409pt}{0.400pt}}
\put(126.0,443.0){\rule[-0.200pt]{2.409pt}{0.400pt}}
\put(935.0,443.0){\rule[-0.200pt]{2.409pt}{0.400pt}}
\put(126.0,474.0){\rule[-0.200pt]{2.409pt}{0.400pt}}
\put(935.0,474.0){\rule[-0.200pt]{2.409pt}{0.400pt}}
\put(126.0,505.0){\rule[-0.200pt]{2.409pt}{0.400pt}}
\put(935.0,505.0){\rule[-0.200pt]{2.409pt}{0.400pt}}
\put(126.0,535.0){\rule[-0.200pt]{2.409pt}{0.400pt}}
\put(935.0,535.0){\rule[-0.200pt]{2.409pt}{0.400pt}}
\put(126.0,535.0){\rule[-0.200pt]{4.818pt}{0.400pt}}
\put(925.0,535.0){\rule[-0.200pt]{4.818pt}{0.400pt}}
\put(126.0,566.0){\rule[-0.200pt]{2.409pt}{0.400pt}}
\put(935.0,566.0){\rule[-0.200pt]{2.409pt}{0.400pt}}

\put(108, 74){\makebox(0,0)[r]{0.85}}
\put(108,228){\makebox(0,0)[r]{0.90}}
\put(108,382){\makebox(0,0)[r]{0.95}}
\put(108,535){\makebox(0,0)[r]{1.00}}
\put(152,37){\makebox(0,0){ $0$}}
\put(254,37){\makebox(0,0){ $2$}}
\put(356,37){\makebox(0,0){ $4$}}
\put(459,37){\makebox(0,0){ $6$}}
\put(561,37){\makebox(0,0){ $8$}}
\put(663,37){\makebox(0,0){$10$}}
\put(766,37){\makebox(0,0){$12$}}
\put(868,37){\makebox(0,0){$14$}}
\put(990,37){\makebox(0,0){$S$}}

\put(152.0,74.0){\rule[-0.200pt]{0.400pt}{4.818pt}}
\put(152.0,546.0){\rule[-0.200pt]{0.400pt}{4.818pt}}
\put(203.0, 74.0){\rule[-0.200pt]{0.400pt}{2.409pt}}
\put(203.0,556.0){\rule[-0.200pt]{0.400pt}{2.409pt}}
\put(254.0, 74.0){\rule[-0.200pt]{0.400pt}{4.818pt}}
\put(254.0,546.0){\rule[-0.200pt]{0.400pt}{4.818pt}}
\put(305.0, 74.0){\rule[-0.200pt]{0.400pt}{2.409pt}}
\put(305.0,556.0){\rule[-0.200pt]{0.400pt}{2.409pt}}
\put(356.0, 74.0){\rule[-0.200pt]{0.400pt}{4.818pt}}
\put(356.0,546.0){\rule[-0.200pt]{0.400pt}{4.818pt}}
\put(408.0, 74.0){\rule[-0.200pt]{0.400pt}{2.409pt}}
\put(408.0,556.0){\rule[-0.200pt]{0.400pt}{2.409pt}}
\put(459.0, 74.0){\rule[-0.200pt]{0.400pt}{4.818pt}}
\put(459.0,546.0){\rule[-0.200pt]{0.400pt}{4.818pt}}
\put(510.0, 74.0){\rule[-0.200pt]{0.400pt}{2.409pt}}
\put(510.0,556.0){\rule[-0.200pt]{0.400pt}{2.409pt}}
\put(561.0, 74.0){\rule[-0.200pt]{0.400pt}{4.818pt}}
\put(561.0,546.0){\rule[-0.200pt]{0.400pt}{4.818pt}}
\put(612.0, 74.0){\rule[-0.200pt]{0.400pt}{2.409pt}}
\put(612.0,556.0){\rule[-0.200pt]{0.400pt}{2.409pt}}
\put(663.0, 74.0){\rule[-0.200pt]{0.400pt}{4.818pt}}
\put(663.0,546.0){\rule[-0.200pt]{0.400pt}{4.818pt}}
\put(715.0, 74.0){\rule[-0.200pt]{0.400pt}{2.409pt}}
\put(715.0,556.0){\rule[-0.200pt]{0.400pt}{2.409pt}}
\put(766.0, 74.0){\rule[-0.200pt]{0.400pt}{4.818pt}}
\put(766.0,546.0){\rule[-0.200pt]{0.400pt}{4.818pt}}
\put(817.0, 74.0){\rule[-0.200pt]{0.400pt}{2.409pt}}
\put(817.0,556.0){\rule[-0.200pt]{0.400pt}{2.409pt}}
\put(868.0, 74.0){\rule[-0.200pt]{0.400pt}{4.818pt}}
\put(868.0,546.0){\rule[-0.200pt]{0.400pt}{4.818pt}}
\put(919.0, 74.0){\rule[-0.200pt]{0.400pt}{2.409pt}}
\put(919.0,556.0){\rule[-0.200pt]{0.400pt}{2.409pt}}

\put(126.0, 74.0){\rule[-0.200pt]{197.297pt}{0.400pt}}
\put(945.0, 74.0){\rule[-0.200pt]{0.400pt}{118.523pt}}
\put(126.0, 74.0){\rule[-0.200pt]{0.400pt}{118.523pt}}
\put(126.0,566.0){\rule[-0.200pt]{197.297pt}{0.400pt}}
\put(945.0, 74.0){\vector(1,0){50}}

\put(152,113){\raisebox{-.8pt}{\makebox(0,0){$\Diamond$}}}
\put(203,119){\raisebox{-.8pt}{\makebox(0,0){$\Diamond$}}}
\put(254,132){\raisebox{-.8pt}{\makebox(0,0){$\Diamond$}}}
\put(305,151){\raisebox{-.8pt}{\makebox(0,0){$\Diamond$}}}
\put(356,177){\raisebox{-.8pt}{\makebox(0,0){$\Diamond$}}}
\put(408,207){\raisebox{-.8pt}{\makebox(0,0){$\Diamond$}}}
\put(459,241){\raisebox{-.8pt}{\makebox(0,0){$\Diamond$}}}
\put(510,278){\raisebox{-.8pt}{\makebox(0,0){$\Diamond$}}}
\put(561,316){\raisebox{-.8pt}{\makebox(0,0){$\Diamond$}}}
\put(612,354){\raisebox{-.8pt}{\makebox(0,0){$\Diamond$}}}
\put(663,393){\raisebox{-.8pt}{\makebox(0,0){$\Diamond$}}}
\put(715,431){\raisebox{-.8pt}{\makebox(0,0){$\Diamond$}}}
\put(766,469){\raisebox{-.8pt}{\makebox(0,0){$\Diamond$}}}
\put(817,504){\raisebox{-.8pt}{\makebox(0,0){$\Diamond$}}}
\put(868,535){\raisebox{-.8pt}{\makebox(0,0){$\Diamond$}}}
\put(919,535){\raisebox{-.8pt}{\makebox(0,0){$\Diamond$}}}
\put(139.00,110.50){\usebox{\plotpoint}}
\put(159.00,112.00){\usebox{\plotpoint}}
\put(179.45,115.43){\usebox{\plotpoint}}
\put(199.86,118.98){\usebox{\plotpoint}}
\put(220.05,123.76){\usebox{\plotpoint}}
\put(240.03,129.26){\usebox{\plotpoint}}
\put(259.76,135.59){\usebox{\plotpoint}}
\put(279.29,142.61){\usebox{\plotpoint}}
\put(298.54,150.35){\usebox{\plotpoint}}
\put(317.53,158.68){\usebox{\plotpoint}}
\put(336.42,167.14){\usebox{\plotpoint}}
\put(354.86,176.60){\usebox{\plotpoint}}
\put(373.10,186.44){\usebox{\plotpoint}}
\put(390.97,196.98){\usebox{\plotpoint}}
\put(408.95,207.30){\usebox{\plotpoint}}
\put(426.41,218.51){\usebox{\plotpoint}}
\put(443.64,230.02){\usebox{\plotpoint}}
\put(460.92,241.45){\usebox{\plotpoint}}
\put(477.95,253.30){\usebox{\plotpoint}}
\put(494.71,265.53){\usebox{\plotpoint}}
\put(511.66,277.50){\usebox{\plotpoint}}
\put(528.27,289.95){\usebox{\plotpoint}}
\put(544.91,302.29){\usebox{\plotpoint}}
\put(561.32,314.99){\usebox{\plotpoint}}
\put(577.80,327.60){\usebox{\plotpoint}}
\put(594.33,340.15){\usebox{\plotpoint}}
\put(610.60,353.02){\usebox{\plotpoint}}
\put(627.19,365.46){\usebox{\plotpoint}}
\put(643.43,378.33){\usebox{\plotpoint}}
\put(659.72,391.14){\usebox{\plotpoint}}
\put(676.22,403.69){\usebox{\plotpoint}}
\put(692.93,415.95){\usebox{\plotpoint}}
\put(709.54,428.40){\usebox{\plotpoint}}
\put(726.02,441.02){\usebox{\plotpoint}}
\put(742.62,453.47){\usebox{\plotpoint}}
\put(759.99,464.74){\usebox{\plotpoint}}
\put(776.82,476.88){\usebox{\plotpoint}}
\put(794.00,488.50){\usebox{\plotpoint}}
\put(811.34,499.89){\usebox{\plotpoint}}
\put(828.91,510.94){\usebox{\plotpoint}}
\put(846.78,521.49){\usebox{\plotpoint}}
\put(864.45,532.36){\usebox{\plotpoint}}
\put(882.67,542.29){\usebox{\plotpoint}}
\put(900.86,552.25){\usebox{\plotpoint}}
\put(919.35,561.67){\usebox{\plotpoint}}
\end{picture}
\caption{The global N\'eel probability $p^*_{\tN}$ versus the total spin $S$ (diamonds) 
and the best-fit line (see text for details).\label{pnspin}}
\end{figure}
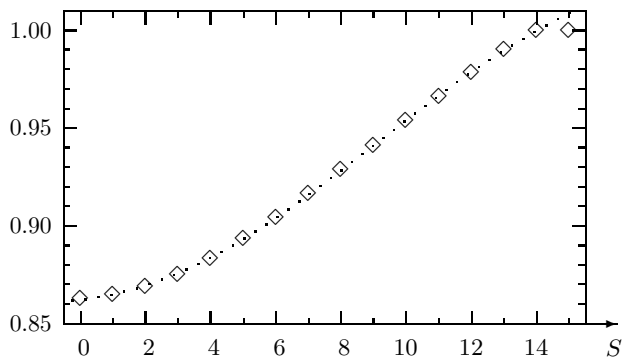

The small ferric wheel Fe$_6$ has been recently investigated by many authors, so we 
have decided to analyze some less discussed properties---spin correlations and 
probability of finding the system in the N\'eel state. In both cases we obtain results suggesting
quite well ordered ground state with highly correlated spins. For example, $\omega^z_1=-2.441$,
$\omega^z_2=1.959$, and $\omega^z_3=-1.952$.\cite{flonewx} It should be stressed that 
$z$-correlations reach maximum magnitude 
in states with $S=1$ (cf.\ Fig.~\ref{corrspin}); the values are $-4.172$, $3.907$,
and $-3.897$, respectively. However, these states are not isotropic, so $x$- and $y$-correlations
are much smaller, whereas total spin correlations $\omega_r=3\omega^z_r$ for $S=0$.
As regards the N\'eel probability, deviation from the classical ordering in the ground state 
can be observed. On the other hand, $p^*_{\tN}=1$ in the cases $S=14,15$ shows that these
multiplets behave classically. It has to be stressed that states with $M=0$ are considered
only. Hence, we obtain energies for all states in a given $S$-multiplet (for $\Gamma=A_1,B_1$).
However, correlations calculated are related to states $\ket{S,0}$---they cannot be
used for states with $M\ne0$. For example, in the case of the unique multiplet $S=15$ 
we have $\omega^z_r=6.25$, $r=1,2,3$, for $M=\pm15$, whereas for $M=0$ we have
$\omega_r^z=-0.2155$, $r=1,2,3$. Plotting $p^*_{\tN}$ versus the total spin $S$ one can
observed almost linear dependence for high $S$ (omitting point $S=15$). A better 
approximation
has been obtained for a function $p^*_{\tN}(S)=\exp({-0.31S}+2.84\sqrt{S}-8.25)+0.86$ (cf.\
Eq.~(\ref{cgc}) and Fig.~\ref{pnspin}). Having determined such relations 
(see also Figs.~\ref{abspin}, 
\ref{corrspin}, \ref{npabx}) for ferric dimers Fe$_2$ and the ferric wheel Fe$_{10}$ we would 
try to approximate results for Fe$_{18}$. The ring Fe$_8$ recently studied by Waldmann et 
al.\cite{wald01} may lead to inconsistent results, since the ground state corresponds to
$\Gamma=A_1$ (see Ref.~\onlinecite{wald01} and  the end of Sec.~\ref{class}). 
Note that there is lack of Fe$_{14}$, but nevertheless it seems interesting to investigate
such a molecule. There are about $8\cdot10^{10}$ basic states, but `only' about $5\cdot10^9$
of them have $M=0$; the total spin $S=0$ leads to 56\,267\,133 states (there are not
basic states, however). Taking into account symmetry properties will decrease these numbers
significantly.   

The methods proposed by Bonner and
Fisher\cite{BF64} almost forty years ago have been
mainly used to estimate properties of magnetic materials in the thermodynamic
limit. On the other hand, magnetic nanoclusters have given rise to more detailed 
investigations of small spin systems comprising ions with high spin number $s$. 
The numeric complexity is similar: there are 60\,466\,176 basic states in the case
of ten spin $s={5\over2}$ what is comparable with that number for 26 spins $s={1\over2}$.
However, our aims are a bit different in these cases. Trying to determine properties
in the thermodynamic limit we observe their dependencies on number of spins $n$, whereas
investigating nanomagnets we rather try to determine a `best-fit' model of a given
molecule. Therefore, in the first case the considerations are mainly limited to one, sometimes
simplified, model, while fitting to experimental data we may need to investigate some different
and rather complicated models.

Amongst many magnetic molecules those containing Fe$^{3+}$
ions seem to be very interesting, since they can be used as models of such metalloprotein 
as ferritin.\cite{stpie96} A fact that this molecule contains as many as 4500 Fe ions 
does not depreciate validity of results obtained for molecules comprising several 
magnetic ions. Experimental data show that noncompensate spin in ferritin is about 50
and it originates from surface randomness.\cite{tejada97} Moreover, these data suggest that 
ferritin exhibits quantum tunneling similar to that observed in Mn$_{12}$ molecules. 

Since recently some papers using a similar approach have been 
published\cite{wald00,wald01,raghu01} we would like to stress differences and advantages
of the method applied in this paper. At first, we go much deeper into structure of 
the permutation representations introducing additional indices (ordered and nonordered
partitions) for basic states. This also leads to some simplifications in the procedure
generating operator matrices.\cite{cpc01,cmstm}  A general formula for matrix 
elements\cite{cpc01,app1} significantly reduces a number of cases to be considered.
Note that this formula needs introduction of double cosets, a structure not mentioned in
the earlier papers. Moreover, many results for the dihedral groups can be presented
in easy-to-use analytical form.\cite{app2} We do not need to use any projection procedure
to derive irreducible basis, because it is determined for transitive representations of 
the dihedral groups in a general form.\cite{app2,lullul} Using any package for multiple
precision arithmetic, like {\sf GMP},\cite{gmp} we are able to find {\it exact}\/ 
eigenvectors of ${\mathbf S}^2$ operator in the case of one-dimensional irreps (and also
for two-dimensional irreps in some simple cases, e.g.\ for D$_6$ or D$_8$, when
$\cos(2\pi/n)$ can be expressed by square roots of integers). This allows determination
of spin correlations and N\'eel probabilities with high accuracy.

\begin{acknowledgments}
This work was supported in part by the State Committee for Scientific Research (KBN)
under the grant No.~2~P03B~074~19. One of us (WF) is indebted to Prof.\ A.~Kerber 
for his hospitality and inspiring discussions during author's stays in Bayreuth (Germany).
Valuable remarks of Dr.\ M.~Thomas are gratefully acknowledged. Some numerical calculations
were carried out in the Supercomputing and Networking Center in Pozna\'n.
\end{acknowledgments}

\appendix

\section{An example}
To illustrate the method for generating operator matrices the case of four spin $s={3\over2}$
is presented below. The solutions of eigenproblems of ${\mathbf S}^2$ and ${\mathcal H}$
as well as discussion of the spin correlations and N\'eel probability are left out, since
at the actual stage of the project they are performed by means of (more or less) standard
numeric methods (the only change is application of procedures from the {\sf GMP} library).

Assuming the isotropic Heisenberg Hamiltonian both $S$ and $M$ are good quantum numbers. 
There are $4^4=256$ Ising configurations and the maximum total spin
number is $S=6$; hence, the possible magnetizations are $M=0,\pm1,\dots,\pm6$. The 
dimensions of subspaces $L_M$ containing states with a given $M$ are as follows:
 \[
   \begin{array}{c|*{7}{r}}
             M & \pm6 & \pm5 & \pm4 & \pm3 & \pm2 & \pm1 &  0 \\ \hline
      \dim L_M &    1 &    4 &   10 &   20 &   31 &   40 & 44 
   \end{array}
 \]  
 It allows determination of numbers $n(S)$ of $S$-multiplets for each $S$ as 
$n(S)=\dim L_{M=S}-\dim L_{M=S+1}$ for $S<6$ and $n(S=6)=1$;\cite{app1} hence one obtains:
 \[
   \begin{array}{c|*{7}{r}}
         S & 6 & 5 & 4 &  3 &  2 & 1 & 0 \\ \hline
      n(S) & 1 & 3 & 6 & 10 & 11 & 9 & 4 
   \end{array}
 \] 
 Of course we have $\sum_{S=0}^6 (2S+1)n(S)=256$ and $\sum_{S=0}^6 n(S)=44=\dim L_{M=0}$
since each $S$-multiplet contains the state $\ket{S\,M=0}$. To investigate nonmagnetic
states with $M=0$ one has to decompose the 44-dimensional space $L_0$ into seven subspaces
labeled by $S$ and with dimensions given in the table presented above. 

To begin with we start from considering {\em ordered}\/ partitions of $n=4$ into no more than
four parts. There are five such partitions, namely
 \[ [4]\,,\; [3,1]\,,\; [2,2]\,,\; [2,1,1]\,,\; [1,1,1,1]\,.  \]
 Each of this partition represents a class of {\em nonordered}\/ partitions of $n=4$
into four non-negative parts $k_0,k_1,k_2$, and $k_3$. There are 35 such partitions collected
in Table~\ref{nonord}. From this table one can learn that there are five nonordered
partitions with $\sum_{l=0}^3 lk_l=6$, i.e.\ the magnetization $M=0$ 
(see Eq.~(\ref{part})). These five partition come in three types: $[3,1]$, $[2,2]$, and 
$[1,1,1,1]$. Therefore, to investigate states with $M=0$ one has to consider only these
three ordered partitions. The number of configurations of type $[k]$ is determined
by the polynomial coefficient $n!/k_0!k_1!k_2!k_3!$ and is the same for all nonordered partitions
of a given type $[\kappa]$. In the case considered, i.e.\ for $[\kappa]=[3,1],[2,2]$,
and $[1,1,1,1]$ these numbers are 4, 6, and 24, respectively, then there are four states
corresponding to $[k]=[0,3,0,1],[1,0,3,0]$, six states for $[k]=[2,0,0,2],[0,2,2,0]$, and
24 states for $[k]=[1,1,1,1]$; 44 states in total. For example, the partition $[1,0,3,0]$
labels a four-element set of states:
 \[
  \{\ket{\textstyle{-{3\over2}},\textstyle{1\over2},\textstyle{1\over2},\textstyle{1\over2}},\;
  \ket{\textstyle{1\over2},\textstyle{-{3\over2}},\textstyle{1\over2},\textstyle{1\over2}},
  \ket{\textstyle{1\over2},\textstyle{1\over2},\textstyle{-{3\over2}},\textstyle{1\over2}},
  \ket{\textstyle{1\over2},\textstyle{1\over2},\textstyle{1\over2},\textstyle{-{3\over2}}}\}
 \]
 since relations $k_0=1$, $k_2=3$, and $k_1=k_3=0$ means that there are one spin
projection ${-{3\over2}}$ and three projections ${1\over2}$. On the other hand, the unique
orbit labeled by $[k]=[1,1,1,1]$ contains  $4!=24$ states with all projections different:
$\ket{{-{3\over2}},{-{1\over2}},{1\over2},{3\over2}},
\ket{{-{1\over2}},{-{3\over2}},{1\over2},{3\over2}}$ etc.

\begin{table}
\caption{Classes of nonordered partitions $[k]$ labeled by ordered partitions $[\kappa]$
for $n=4$ and $s={3\over2}$; in the third column the magnetization $M=\sum_{l=0}^3 lk_l-6$ 
is calculated. \label{nonord}}

\begin{ruledtabular}
\begin{tabular}{ccrc}
     $[\kappa]$ & $[k]$ & \multicolumn{1}{c}{$M$} & orbit representative \\ \hline
 [4]       & [4,0,0,0] &--6 & $\ket{-3/2,-3/2,-3/2,-3/2}$ \\ 
           & [0,4,0,0] &--2 & $\ket{-1/2,-1/2,-1/2,-1/2}$ \\  
           & [0,0,4,0] &  2 & $\ket{+1/2,+1/2,+1/2,+1/2}$ \\ 
           & [0,0,0,4] &  6 & $\ket{+3/2,+3/2,+3/2,+3/2}$ \\ \hline
 [3,1]     & [3,1,0,0] &--5 & $\ket{-3/2,-3/2,-3/2,-1/2}$ \\
           & [3,0,1,0] &--4 & $\ket{-3/2,-3/2,-3/2,+1/2}$ \\
           & [3,0,0,1] &--3 & $\ket{-3/2,-3/2,-3/2,+3/2}$ \\
           & [1,3,0,0] &--3 & $\ket{-3/2,-1/2,-1/2,-1/2}$ \\
           & [0,3,1,0] &--1 & $\ket{-1/2,-1/2,-1/2,+1/2}$ \\
           & [0,3,0,1] &  0 & $\ket{-1/2,-1/2,-1/2,+3/2}$ \\
           & [1,0,3,0] &  0 & $\ket{-3/2,+1/2,+1/2,+1/2}$ \\
           & [0,1,3,0] &  1 & $\ket{-1/2,+1/2,+1/2,+1/2}$ \\
           & [0,0,3,1] &  3 & $\ket{+1/2,+1/2,+1/2,+3/2}$ \\
           & [1,0,0,3] &  3 & $\ket{-3/2,+3/2,+3/2,+3/2}$ \\
           & [0,1,0,3] &  4 & $\ket{-1/2,+3/2,+3/2,+3/2}$ \\
           & [0,0,1,3] &  5 & $\ket{+1/2,+3/2,+3/2,+3/2}$ \\ \hline
 [2,2]     & [2,2,0,0] &--4 & $\ket{-3/2,-3/2,-1/2,-1/2}$ \\
           & [2,0,2,0] &--2 & $\ket{-3/2,-3/2,+1/2,+1/2}$ \\
           & [2,0,0,2] &  0 & $\ket{-3/2,-3/2,+3/2,+3/2}$ \\
           & [0,2,2,0] &  0 & $\ket{-1/2,-1/2,+1/2,+1/2}$ \\ 
           & [0,2,0,2] &  2 & $\ket{-1/2,-1/2,+3/2,+3/2}$ \\
           & [0,0,2,2] &  4 & $\ket{+1/2,+1/2,+3/2,+3/2}$ \\ \hline
 [2,1,1]   & [2,1,1,0] &--3 & $\ket{-3/2,-3/2,-1/2,+1/2}$ \\
           & [2,1,0,1] &--2 & $\ket{-3/2,-3/2,-1/2,+3/2}$ \\
           & [2,0,1,1] &--1 & $\ket{-3/2,-3/2,+1/2,+3/2}$ \\
           & [1,2,1,0] &--2 & $\ket{-3/2,-1/2,-1/2,+1/2}$ \\ 
           & [1,2,0,1] &--1 & $\ket{-3/2,-1/2,-1/2,+3/2}$ \\
           & [0,2,1,1] &  1 & $\ket{-1/2,-1/2,+1/2,+3/2}$ \\
           & [1,1,2,0] &--1 & $\ket{-3/2,-1/2,+1/2,+1/2}$ \\
           & [1,0,2,1] &  1 & $\ket{-3/2,+1/2,+1/2,+3/2}$ \\
           & [0,1,2,1] &  2 & $\ket{-1/2,+1/2,+1/2,+3/2}$ \\
           & [1,1,0,2] &  1 & $\ket{-3/2,-1/2,+3/2,+3/2}$ \\
           & [1,0,1,2] &  2 & $\ket{-3/2,+1/2,+3/2,+3/2}$ \\
           & [0,1,1,2] &  3 & $\ket{-1/2,+1/2,+3/2,+3/2}$ \\ \hline
 [1,1,1,1] & [1,1,1,1] &  0 & $\ket{-3/2,-1/2,+1/2,+3/2}$     
\end{tabular}
\end{ruledtabular}
\end{table}

\begin{figure}[t]
 \begin{picture}(120,120)(-20,-20)
 \put(0,0){\framebox(80,80){}}
 \put(  0, -5){\makebox(0,0)[tl]{4}}
 \put( 80, -5){\makebox(0,0)[tr]{1}}
 \put( 80, 85){\makebox(0,0)[br]{2}}
 \put(  0, 85){\makebox(0,0)[bl]{3}}
 
 \put(40,40){\makebox(0,0){\large$\blacksquare$}}
 \put( 85, -3){\makebox(0,0)[bl]{$U_0$}}
 \put( 85, 35){\makebox(0,0)[tl]{$U_1$}}
 \put( 85, 80){\makebox(0,0)[tl]{$U_2$}}
 \put( 45, 85){\makebox(0,0)[bl]{$U_3$}}
 \put(0,0){\circle*{6}}
 \put(80,0){\circle*{6}}
 \put(0,80){\circle*{6}}
 \put(80,80){\circle*{6}}

 \thicklines

 \put(-20,40){\line(1,0){120}}
 \put(40,-20){\line(0,1){120}}
 \put(-10,-10){\line( 1,1){100}}
 \put( 90,-10){\line(-1,1){100}}
 \end{picture}
 \caption{Labeling scheme of elements in the dihedral group $\mathrm{D}_4$ and nodes
(spins) for a system of four spins. A black square denotes the four-fold axis and thick lines 
denote two-fold axes.}\label{d4}
\end{figure}
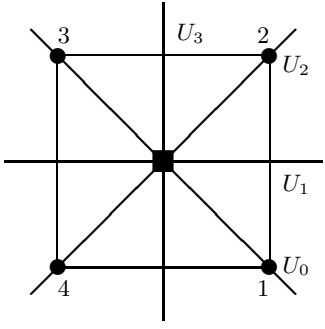

 The decomposition of an orbit $O[k]$ (of the symmetric group $\Sigma_4$) into orbits
of a Hamiltonian symmetry group $G$ (isomorphic to the dihedral group $\mathrm{D}_4$ in
the case considered) depends on its type, i.e.\ on the ordered partition $[\kappa]$.
Before presenting these decompositions a brief outlook of the group $\mathrm{D}_4$ is 
necessary. This group is generated by the four-fold rotation $C_4$ and the two-fold
rotation $U_0$ with the generating relations
 \[
   C_4^4=E\,,\; U_0^2=E\,,\; (C_4U_0)^2=U_1^2=E\,,
 \]
 where $E$ denotes the unit element in $\mathrm{D}_4$. The labeling scheme of group
elements and numbering of nodes (spins) is presented in Fig.~\ref{d4}. There are five
classes of conjugated elements: $\{E\}$, $\{C_2\}$, $\{C_4,C_4^{-1}\}$, $\{U_0,U_2\}$,
$\{U_1,U_3\}$ and, hence, five irreducible representations: $A_1$, $A_2$, $B_1$, $B_2$,
$E_1$ with characters presented in Table~\ref{char}. The dihedral group $\mathrm{D}_4$
has ten subgroups collected in eight classes of conjugated subgroups (in this very 
simple case only two classes contain two subgroups). However, three subgroups 
are excluded from a set of possible configuration stabilizers.\cite{ker,app2} It results in five
classes of subgroups or, in the other words, five types of orbits. These classes
are represented by subgroups:
 \begin{eqnarray*}
 \mathrm{C}_1&=&\{E\}\,,\quad \mathrm{D}_1^0\;=\;\{E,U_0\}\,,\quad 
        \mathrm{D}_1^1\;=\;\{E,U_1\}\,,\\
 \mathrm{D}_2^0&=&\{E,C_2,U_0,U_2\}\,,\\ 
   \mathrm{D}_4&=&
\{E,C_4,C_2,C_4^{-1},U_0,U_1,U_2,U_3\}\,.
 \end{eqnarray*}

\begin{table}
\caption{Characters of irreducible representations of the dihedral group 
${\mathrm D}_6$.  \label{char}}

\begin{ruledtabular}
\begin{tabular}{c|*{5}{c}}
$\Gamma$ & $\{E\}$ & $\{C_4,C_4^{-1}\}$ & $\{C_2\}$ &  $\{U_0,U_2\}$ 
   & $\{U_1,U_3\}$ \\ \hline
$A_1$ & 1 &   1 &   1 &   1 &   1 \\
$A_2$ & 1 &   1 &   1 & --1 & --1 \\
$B_1$ & 1 & --1 &   1 & --1 &   1 \\
$B_2$ & 1 & --1 &   1 &   1 & --1 \\
$E_1$ & 2 &   0 & --2 &   0 &   0
\end{tabular}
\end{ruledtabular}
\end{table}

The four-element orbit labeled by the partition $[3,1]$ is also an orbit of the
dihedral group $\mathrm{D}_4$ with a stabilizer $\mathrm{D}_1^0$. As representatives
of orbits of this type one can choose a state in the form $\ket{a,b,b,b}$. In the case
considered here, two orbits of this type are represented by states
$\ket{{3\over2},{-{1\over2}},{-{1\over2}},{-{1\over2}}}$ and 
$\ket{{-{3\over2}},{1\over2},{1\over2},{1\over2}}$, respectively. The orbits of the
type $[2,2]$ have been considered in the main text. Each of them decomposes into two orbits
of the group $\mathrm{D}_4$: the first has a stabilizer $\mathrm{D}_1^1$, whereas 
the second one has a stabilizer $\mathrm{D}_2^0$. In this way one obtains four 
new orbit representatives:
 \begin{eqnarray*}
  {}[2,0,0,2]\,\colon&& \ket{\textstyle{-{3\over2}},\textstyle{-{3\over2}},
     \textstyle{3\over2},\textstyle{3\over2}}\,,\;
 \ket{{-\textstyle{3\over2}},\textstyle{3\over2},\textstyle{-{3\over2}},\textstyle{3\over2}}\,;\\
  {}[0,2,2,0]\,\colon&& \ket{\textstyle{-{1\over2}},\textstyle{-{1\over2}},
    \textstyle{1\over2},\textstyle{1\over2}}\,,\;
 \ket{\textstyle{-{1\over2}},\textstyle{1\over2},\textstyle{-{1\over2}},\textstyle{1\over2}}\,.  
 \end{eqnarray*}
 The last, 24-element, orbit decomposes into three orbits in the same type: all of them
have trivial stabilizer $\mathrm{C}_1$, so each of them is the so-called regular orbit
of $\mathrm{D}_4$.\cite{ker} As representatives of these eight-element orbits one can choose
states
 \[
  \ket{\textstyle{-{3\over2}},\textstyle{-{1\over2}},\textstyle{1\over2},\textstyle{3\over2}}\,,\;
  \ket{\textstyle{-{3\over2}},\textstyle{1\over2},\textstyle{3\over2},\textstyle{-{1\over2}}}\,,\;
  \ket{\textstyle{-{3\over2}},\textstyle{3\over2},\textstyle{-{1\over2}},\textstyle{1\over2}}\,.
 \]
In this way the complete group-theoretical (and combinatorial) classification of 
all 44 Ising configurations with $M=0$ has been done. For example the state 
$\mu=\ket{{1\over2},{-{1\over2}},{1\over2},{-{1\over2}}}$ is labeled by 
the nonordered partition
$[k]=[0,2,2,0]$ (so the ordered partition is $[2,2]$); its stabilizer is $\mathrm{D}_2^0$ so
it belongs to the orbit represented by $\nu=\ket{{-{1\over2}},{1\over2},{-{1\over2}},{1\over2}}$ 
and as a representative $g_r$ of the left coset $g_r\mathrm{D}_2^0\in\mathrm{D}_4/\mathrm{D}_2^0$
one can choose the four-fold rotation $C_4$, since $\mu=C_4\nu$.

The partition $[k]$ may be used to simplify determination of matrix elements of ${\mathbf S}^2$
or ${\mathcal H}$. Assuming that only bilinear terms are present one can notice that states
labeled by $[2,0,0,2]$ can be transformed into states labeled by $[1,1,1,1]$ only. States of 
the first type comprise two projections ${3\over2}$ and two projections ${-{3\over2}}$. Therefore,
the only nontrivial action of a bilinear term $s_j^+s_k^-$ is possible when $s_j^+$ acts on the
projection ${-{3\over2}}$ and $s_k^-$ acts on ${3\over2}$. It results in the state containing
all four different projections, i.e.\ of the type $[1,1,1,1]$. Similar discussion of all
five partitions leads to the following graph:\cite{cpc01,cmstm}
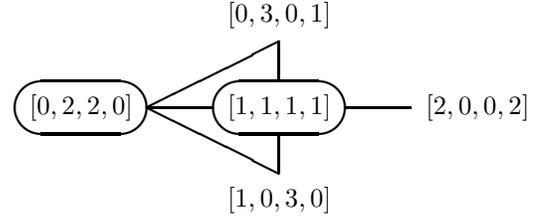
 \begin{center}
  \begin{picture}(195,80)( -5,0)
   \put( 20,40){\makebox(0,0){$[0,2,2,0]$}}
   \put( 95,40){\makebox(0,0){$[1,1,1,1]$}}
   \put( 95, 0){\makebox(0,0)[b]{$[1,0,3,0]$}}
   \put( 95,80){\makebox(0,0)[t]{$[0,3,0,1]$}}
   \put(170,40){\makebox(0,0){$[2,0,0,2]$}}
   \thicklines
   \put(145,40){\line(-1,0){25}}
   \put(45,40){\line(1,0){25}}
   \put(95,15){\line(0,1){15}}
   \put(95,50){\line(0,1){15}}
   \put(45,40){\line(2,1){50}}
   \put(45,40){\line(2,-1){50}}
   \put(95,40){\oval(50,20)}
   \put(20,40){\oval(50,20)}
  \end{picture}
 \end{center}
 The lines join partitions, which are related by the action of bilinear terms $s_j^+s_k^-$;
encircled partitions denote loops in this graph, i.e.\ for each state labeled by these 
partitions exist such a term $s_j^+s_k^-$ that a state labeled by the same partition is 
obtained after its action. For example,
$s_1^+s_2^-\ket{{-{1\over2}},{1\over2},{1\over2},{-{1\over2}}}= 
4\ket{{1\over2},{-{1\over2}},{1\over2},{-{1\over2}}}$. 

The obtained orbits have four different stabilizers: $\mathrm{C}_1$, $\mathrm{D}_1^0$, 
$\mathrm{D}_1^1$, and $\mathrm{D}_2^0$. Note that there are three orbits of the first type
and two orbits of each of the other types. To introduce the linear structure one has to 
decompose (transitive) permutation representation $R^{\mathrm{D}_4:U}$ into (linear) 
irreducible ones.\cite{app1,app2} For these four groups these decompositions are as follows:
 \begin{eqnarray*} 
   R^{\mathrm{D}_4:\mathrm{C}_1}  &=& A_1\oplus A_2\oplus B_1\oplus B_2\oplus 2E_1\,,\\ 
   R^{\mathrm{D}_4:\mathrm{D}_1^0}&=& A_1\oplus B_1\oplus E_1\,,\\ 
   R^{\mathrm{D}_4:\mathrm{D}_1^1}&=& A_1\oplus B_2\oplus E_1\,,\\ 
   R^{\mathrm{D}_4:\mathrm{D}_2^0}&=& A_1\oplus B_1\,. 
 \end{eqnarray*} 
 Therefore, the 44-dimensional space $L_0$ can be decomposed into nine subspaces labeled
by $\Gamma=A_1$, three by $\Gamma=A_2$, seven by $\Gamma=B_1$, five by $\Gamma=B_2$,  and
ten by $\Gamma=E_1$. Analogous considerations for 40-dimensional space $L_{M=1}$ give us 
numbers 7, 3, 7, 3, and 10, respectively. It means that amongst four ($44-40$) multiplets 
with $S=0$ two ($9-7$) are labeled by $A_1$, and the other two ($5-3$) by $B_2$. In the case of
antiferromagnetic interactions the ground
state has the symmetry $A_1$, so it belongs to the first pair.


\begin{thebibliography}{52}

\bibitem{can91} A.~Caneschi, D.~Gatteschi, R.~Sessoli, A.-L.~Barra, L.~C.~Brunel,
 and M.~Guillot, J.~Am.\ Chem.\ Soc. {\bf113}, 5873 (1991).

\bibitem{nat93} R.~Sessoli, D.~Gatteschi, A.~Caneschi, and M.~A. Novak, Nature (London) 
 {\bf365}, 141 (1993).
 
\bibitem{lis} T.~Lis, Acta Crystallogr., Sect.\ B: Struct.\ Crystallogr.\ Cryst.\ Chem.
 {\bf36}, 2042 (1980).

\bibitem{can88} A.~Caneschi, D.~Gatteschi, J.~Laugier, P.~Rey, R.~Sessoli, and C.~Zanchini,
 J. Am.\ Chem.\ Soc. {\bf110}, 2795 (1988).

\bibitem{jpcm01} A.~Caramico~D'Auria, U.~Esposito,  E.~Esposito, G.~Kamieniarz, and 
 R.~Matysiak, J. Phys.: Condens.\ Matter {\bf13}, 2017 (2001).

\bibitem{las98} A.~Lascialfari,  D.~Gatteschi, A.~Cornia, U.~Balucani, M.~G. Pini, and
 A.~Rettori, Phys.\ Rev.\ B {\bf57}, 1115 (1998).

\bibitem{scu96} A.~Scuiller, T.~Mallah, M.~Verdaguer, A.~Nivorozkhin, J.-L. Tholence, and
 P.~Veillet, New J.\ Chem. {\bf20}, 1 (1996).

\bibitem{blake} A.~J. Blake, C.~M. Grant, S.~Parsons, J.~M. Rawson, and R.~E.~P. Winpenny, 
 J. Chem.\ Soc., Chem.\ Commun. {\bf20}, 2363 (1994).

\bibitem{fe02} A.~Lascialfari, F.~Tabak, G.~L. Abbati,  F.~Borsa, M.~Corti, and D.~Gatteschi,
 J. Appl.\ Phys. {\bf85}, 4539 (1999).

\bibitem{can99} A.~Caneschi, D.~Gatteschi, C.~Sangregorio, R.~Sessoli, L.~Sorace,
 A.~Cornia, M.~A. Novak, C.~Paulsen, and W.~Wernsdorfer, J. Magn.\ Magn.\ Mater. {\bf200},
 182 (1999).

\bibitem{papa92} G.~C. Papaefthymiou, Phys.\ Rev.\ B {\bf46}, 10366 (1992).

\bibitem{wern00} I.~Chiorescu, W.~Wernsdorfer, A.~M\"uller, H.~B\"ogge, and B.~Barbara,
 Phys.\ Rev.\ Lett. {\bf84}, 3454 (2000).

\bibitem{jamet00} M.~Jamet, V.~Dupuis, P.~M\'elinon, G.~Guiraud, A.~P\'erez, W.~Wernsdorfer,
 A.~Traverse, and B.~Baguenard, Phys.\ Rev.\ B {\bf62}, 493 (2000).

\bibitem{jamet01} M.~Jamet, W.~Wernsdorfer, C.~Thirion, D.~Mailly, V.~Dupuis, P.~M\'elinon,
 and A.~P\'erez, Phys.\ Rev.\ Lett. {\bf86}, 4676 (2001).

\bibitem{blon90} G.~Blondin and J.~J. Giererd, Chem.\ Rev. {\bf90}, 1359 (1990).

\bibitem{stpie96} T.~G. St.~Pierre, P.~Chan, K.~R. Bauchspiess, J.~Webb, S.~Betteridge,
 S.~Walton, and D.~P.~E. Dickson, Coord.\ Chem.\ Rev. {\bf151}, 125 (1996).

\bibitem{tejada97} J.~Tejada, X.~X. Zhang, E.~del Barco, J.~M. Hern\'andez, and
 E.~M. Chudnovsky, Phys.\ Rev.\ Lett. {\bf79}, 1754 (1997).

\bibitem{sell99} D.~J. Sellmyer, M.~Yu, and R.~D. Kirby, Nanostruct.\ Mater. {\bf12},
 1021 (1999).

\bibitem{fe08b} A.-L. Barra, P.~Debrunner, D.~Gatteschi, Ch.~E.~Schultz, and R.~Sessoli,
 Europhys.\ Lett. {\bf35}, 133 (1996).

\bibitem{sang97} C.~Sangregorio, T.~Ohm, C.~Paulsen, R.~Sessoli, and D.~Gatteschi,
 Phys.\ Rev.\ Lett. {\bf78}, 4645 (1997).

\bibitem{fe08c} W.~Wernsdorfer, A.~Caneschi, R.~Sessoli, D.~Gatteschi, A.~Cornia, V.~Villar,
 and C.~Paulsen, Phys.\ Rev.\ Lett. {\bf84}, 2965 (2000).

\bibitem{prok98} N.~V. Prokof'ev and P.~C.~E.  Stamp, Phys.\ Rev.\ Lett.  {\bf80}, 5794 (1998).

\bibitem{wald01} O.~Waldmann, R.~Koch, S.~Schromm, J.~Sch\"ulein, P.~M\"uller, I.~Bernt,
 R.~W. Saalfrank, F.~Hampel, and E.~Balthes, Inorg.\ Chem. {\bf40}, 2986 (2001).

\bibitem{saal97} R.~W. Saalfrank, I.~Brent, E.~Uller, and F.~Hampel, Angew.\ Chem. {\bf109},
 2596 (1997).

\bibitem{fe06a} A.~Lascialfari, D.~Gatteschi, F.~Borsa, and A.~Cornia, Phys.\ Rev.\ B {\bf55},
 14341 (1997).

\bibitem{fe06b} M.~Affronte, J.~C. Lasjaunias, A.~Cornia, and A.~Caneschi, Phys.\ Rev.\ B
 {\bf60}, 1161 (1999).

\bibitem{fe06c} A.~Cornia, A.~G.~M. Jansen, and M.~Affronte, Phys.\ Rev.\ B {\bf60},
 12177 (1999).

\bibitem{bengat} A.~Bencini and D.~Gatteschi, {\it EPR of Exchange Coupled Systems}
 (Springer-Verlag, Berlin, 1990).

\bibitem{ker} A.~Kerber, {\it Algebraic Combinatorics via Finite Group Action} (BI
 Wissenshaftsverlag, Mannheim-Wien-Z\"urich, 1991).

\bibitem{jmmm99} G.~Kamieniarz, R.~Matysiak, W.~Florek, and S.~Wa{\l}cerz, 
 J. Magn.\ Magn.\ Mater. {\bf203}, 271 (1999).

\bibitem{cpc01} W.~Florek, Comp.\ Phys.\ Commun. {\bf138}, 264 (2001).

\bibitem{app1} W.~Florek, Acta Phys.\ Polon.\ A {\bf100}, 3 (2001), and
 references therein.

\bibitem{app2} S.~Bucikiewicz, L.~D\c{e}bski, and W.~Florek, Acta Phys.\ Polon.\ A {\bf100},
 453 (2001).

\bibitem{fe06d} B.~Normand, X.~Wang, X.~Zotos, and D.~Loss, Phys.\ Rev.\ B {\bf63},
 184409 (2001).

\bibitem{fe04x} G.~Amoretti, S.~Carretta, R.~Caciuffo, H.~Casalta, A.~Cornia, M.~Affronte, and
 D.~Gatteschi, Phys.\ Rev.\ B {\bf64}, 104403 (2001).

\bibitem{fe08a} K.~Wieghardt, K.~Pohl, I.~Jibril, and G.~Huttner, 
 Angew.\ Chem.\ Int.\ Ed.\ Engl.  {\bf23}, 77 (1984).

\bibitem{fe04} H.~Oshio, N.~Hoshino, and T.~Ito, J.  Am.\ Chem.\ Soc. {\bf122}, 12602 (2000).

\bibitem{fe10} Z.~Zeng, Y.~Duan, and D.~Guenzburger, Phys.\ Rev.\ B {\bf55}, 12522 (1997).

\bibitem{cmsta} W.~Florek, Comp.\ Meth.\ Sci.\ Tech. {\bf7}, 41 (2001).

\bibitem{cmstm} S.~Bucikiewicz and W.~Florek, Comp.\ Meth.\ Sci.\ Tech. {\bf7}, 27 (2001).

\bibitem{BF64} J.~C. Bonner and M.~E. Fisher, Phys.\ Rev. {\bf135}, A640 (1964).

\bibitem{wald00} O.~Waldmann, Phys.\ Rev.\ B {\bf61}, 6138 (2000).

\bibitem{kerjam} A.~Kerber and G.~D. James, {\it The Representation Theory of the Symmetric 
 Group} (Adison--Wesley, Reading, MA, 1981).

\bibitem{lullul} B.~Lulek and T.~Lulek, J.  Phys.\ A: Math.\ Gen. {\bf17}, 3077 (1984), and
 references therein.

\bibitem{marsh} W.~Marshall, Proc.\ Roy.\ Soc.\ (London) {\bf A232}, 48 (1955).

\bibitem{casp} P.~L. Iske and W.~J. Caspers, Physica {\bf142A}, 360 (1987).

\bibitem{lm} E.~Lieb and D.~Mattis, J. Math.\ Phys. {\bf3}, 749 (1962).  

\bibitem{lsm} E.~Lieb, T.~Schultz, and D.~Mattis, Ann.\ Phys.\ (N.Y.) {\bf16}, 407 (1961).

\bibitem{gmp} T.~Granlund, {\it The GNU Multiple Precision Arithmetic Library} (Free Software
 Foundation, Boston, MA, 2000), ver.\ 3.1.1, for up-to-date information on {\sf GMP} see the 
 {\sf GMP} Home Pages at {\tt http://www.swox.se/gmp/}.

\bibitem{flonewx} W.~Florek, J.\ Magn.\ Magn.\ Mater.\ {bf247}, 200 (2002) 
 [scheduled publication].

\bibitem{edm} A.~R. Edmonds, {\it Angular Momentum in Quantum Mechanics} (Princeton Univ.\ Press,
 Princeton, 1957).

\bibitem{raghu01} C.~Raghu, I.~Rudra, D.~Sen, and S.~Ramasesha, Phys.\ Rev.\ B {\bf64},
 0644419 (2001).

\end{thebibliography}
\end{document}